\documentclass[twocolumn, amsmath,amssymb, aps, showpacs, graphicx]{revtex4-2}

\usepackage{graphicx}
\usepackage{dcolumn}
\usepackage{bm}
\usepackage{lineno, hyperref}
\usepackage{natbib}
\begin{document}

\title{Coexistence of superconductivity and excitonic pairing in a doped-biased double-layer system}
\author{V. Apinyan} 
\altaffiliation[e-mail:]{v.apinyan@intibs.pl}
\author{M. Sahakyan} 
\affiliation{Institute of Low Temperature and Structure Research, Polish Academy of Sciences\\ 50-422, Wroc\l{}aw 2, Poland 
}
%
\begin{abstract}
The subject of the present study is the double-layer square-lattice system with the intralayer phonon modulations. We investigate the superconducting and excitonic pairings, as well as their coexistence, as functions of various physical parameters in the system. These parameters include temperature, intralayer and interlayer Coulomb interactions, the electron-phonon coupling parameter, doping and the applied electric field. The existence of superconductivity is demonstrated by considering the influence of intralayer phonons on the total charge density leading to the modification of the total energy of the electrons. Our results provides insights into the long-standing problem of the mechanism of superconductivity in high-$T_c$ cuprate superconductors.          
\end{abstract}     



\maketitle
%
\section{\label{sec:Section_1} Introduction}
%

One of the most intriguing problems in condensed matter physics is the explanation of the mechanism that leads to superconducting pairing in high temperature superconductors (high-T$_C$). Effective t-J theories have successfully explained many fundamental properties of the high-T$_C$ cuprates \cite{cite_1, cite_2, cite_3, cite_4, cite_5, cite_6, cite_7}. An alternative mechanism for anisotropic spin-singlet supercondutivity in high-T$_C$ cuprates has been proposed, demonstrating this near the antiferromagnetic transition edge without involving phonons, based on weak-coupling perturbation theory \cite{cite_8, cite_9, cite_10}. Moreover, within two-dimensional Hubbard model, it has been shown that antiferromagnetic fluctuations serve as the pairing glue for superconductivity in both underdoped and the overdoped regimes, particularly at the intermediate coupling limit \cite{cite_11}. In Ref.\cite{cite_12} it was shown that antiferromagnetic spin fluctuations with finite momentum ${\bf{q}}$ can lead to $p$-wave spin-triplet superconducting instabilities.     

Although these mechanisms shed light on the unconventional nature of superconductivity in high-T$_C$ cuprates, the role of the electron-phonon mechanism in providing a comprehensive description of superconducting states remains a topic of considerable debate \cite{cite_13, cite_14, cite_15}.  

Recently, accurate many body calculations have been conducted in the context of the extended Hubbard model for one-dimensional cuprate chains \cite{cite_16}, demonstrating spin-triplet superconducting pairing by considering attractive nearest-neighbor Coulomb interaction.  
In contrast to the proposed spin-singlet mechanisms \cite{cite_9, cite_17, cite_18, cite_19, cite_20, cite_21} for pairing in high-T$_C$ YBa${_2}$Cu${_3}$O${_{7-x}}$ and YBa$_2$Cu$_3$O$_{7-x}$F$_1$ superconductors, intense zero- field absorption line observed in electronic paramagnetic resonance measurements \cite{cite_22, cite_23} suggests that the pairing in these compounds is more likely spin-triplet rather than spin-singlet.   

In the present work, we study the coexistence of spin-singlet excitonic and spin-triplet superconducting phases within the generalized Hubbard model for a metallic square lattice bilayer, considering next-nearest hopping and electron-phonon interaction within the layers. Using this model we calculate a series of important physical parameters and we present the resulting phase transition diagram. Beyond high-temperature superconductors, our findings could be applied to any bilayer system composed of two metallic square lattices. In our theory, single-particle excitations (with a given mode) are represented as the factorization of two, simultaneous, single-particle excitations (with different modes), as previously proposed by us in another context \cite{cite_24}. The interlayer hopping, intralayer Hubbard interaction and interlayer on-site repulsive Coulomb interactions have been appropriately included in the calculations. Our results show that the principal mechanism for the emergence of the superconducting states is the electron-phonon interaction in the system. We also show the doping dependence of the calculated physical quantities and propose a new phase diagram for the spin-triplet pairing in such systems. The proposed model could serve as a new mechanism for the spin-triplet superconducting states in high-T$_C$ cuprates.     

The paper is organized as follows: In Section \ref{sec:Section_2}, we introduce the Hamiltonian of the system and present the electron-phonon coupling mechanism prior to discussing superconductivity. In Section \ref{sec:Section_3}, we discuss the obtained numerical results, and in the Section \ref{sec:Section_4}, we provide a brief conclusion. In Appendix \ref{sec:Section_5}, we offer additional details about the self-consistent equations derived in this work.         
%
\section{\label{sec:Section_2} The Hamiltonian}
%
\subsection{\label{sec:Section_2_1} Electron density coupling mechanism}
%
The bilayer model considered here consists of two metallic square lattice layers arranged one on top of the other and described by the following Hamiltonian:
\begin{eqnarray}
{\cal{\hat{H}}}={\cal{\hat{H}}}_{0}+{\cal{\hat{H}}}_{\rm int},
\label{Equation_1}
\end{eqnarray}
where ${\cal{\hat{H}}}_{0}$ is the non-interacting part and ${\cal{\hat{H}}}_{\rm int}$ is the interaction Hamiltonian. The non-interacting part is given by:
\begin{eqnarray}
{\cal{\hat{H}}}_{0}&&=-t_{0}\sum_{\left\langle{\bf{r}}{\bf{r}}'\right\rangle}\sum_{\ell=1,2}\sum_{\sigma}\left({\hat{a}}^{\dag}_{\ell\sigma}\left({\bf{r}}\right)\hat{a}_{\ell\sigma}\left({\bf{r}}'\right)+\rm{h.c.}\right)-
\nonumber\\
&&-t_{1}\sum_{\left\langle\left\langle{\bf{r}}{\bf{r}}''\right\rangle\right\rangle}\sum_{\ell=1,2}\sum_{\sigma}\left({\hat{a}}^{\dag}_{\ell\sigma}\left({\bf{r}}\right)\hat{a}_{\ell\sigma}\left({\bf{r}}''\right)+\rm{h.c.}\right)-
\nonumber\\
&&-t_{\perp}\sum_{{\bf{r}},\sigma}\left({\hat{a}}^{\dag}_{1\sigma}\left({\bf{r}}\right)\hat{a}_{2\sigma}\left({\bf{r}}\right)+\rm{h.c.}\right)-
\nonumber\\
&&-\mu\sum_{{\bf{r}}\ell}\hat{n}_{\ell}\left({\bf{r}}\right)
\nonumber\\
&&-\sum_{{\bf{r}}\ell}\varepsilon\left({\bf{r}}\right)\hat{c}^{\dag}_{\ell}\left({\bf{r}}\right)\hat{c}_{\ell}\left({\bf{r}}\right)
\nonumber\\
\label{Equation_2}
\end{eqnarray}
and the interacting part is
\begin{eqnarray}
{\cal{\hat{H}}}_{\rm int}&&=-\alpha\sum_{{\bf{r}}\ell}\hat{n}_{\ell}\left({\bf{r}}\right){\rm{div{\hat{R}}}}_{\ell}\left({\bf{r}}\right)+U\sum_{{\bf{r}}\ell}\hat{n}_{\ell\uparrow}\left({\bf{r}}\right)\hat{n}_{\ell\downarrow}\left({\bf{r}}\right)+
\nonumber\\
&&+\frac{W}{2}\sum_{{\bf{r}}\sigma\sigma'}\left(\hat{n}_{1\sigma}\left({\bf{r}}\right)\hat{n}_{2\sigma'}\left({\bf{r}}\right)+\hat{n}_{2\sigma}\left({\bf{r}}\right)\hat{n}_{1\sigma'}\left({\bf{r}}\right)\right)
\nonumber\\
&&+\frac{V}{2}\sum_{{\bf{r}}}\left(\hat{n}_{2}\left({\bf{r}}\right)-\hat{n}_{1}\left({\bf{r}}\right)\right).
\nonumber\\
\label{Equation_3}
\end{eqnarray}
The first and second terms in Eq.(\ref{Equation_2}) represent the intralayer nearest-neighbor ${\bf{r}}'$ and next-nearest neighbor ${\bf{r}}''$hopping terms, associated with the hopping parameters $t_{0}$ ($t_0=0.3$ eV, according to Ref.\cite{cite_25}) and $t_{1}$, respectively. The third term describes the hopping between the layers, characterized by the hopping parameter $t_{\perp}$, while the fourth term represents the chemical potential. The operators $a^{\dag}_{\ell\sigma}\left({\bf{r}}\right)$ and $a_{\ell\sigma}\left({\bf{r}}\right)$ ($\ell=1,2$) are the electron creation and annihilation operators in the top and bottom layers, with layer indices $\ell=1$ and $\ell=2$, and index $\sigma$ denotes the spin of the electrons: $\sigma=\uparrow,\downarrow$. The operators $\hat{n}_{\ell}\left({\bf{r}}\right)$ represent the electron density in each layer.
\begin{eqnarray}
\hat{n}_{\ell}\left({\bf{r}}\right)=\sum_{\sigma=\uparrow\downarrow}{\hat{a}}^{\dag}_{\ell\sigma}\left({\bf{r}}\right){\hat{a}}_{\ell\sigma}\left({\bf{r}}\right).
\label{Equation_4}
\end{eqnarray}
The last term in Eq.(\ref{Equation_2}) corresponds to free phonons. The operators $\hat{c}^{\dag}\left({\bf{r}}\right)$ and $\hat{c}\left({\bf{r}}\right)$ are the phonon creation and annihilation operators, and $\varepsilon\left({\bf{r}}\right)$ represents the excitation energy of a single phonon quantum. 

The interaction Hamiltonian in Eq.(\ref{Equation_3}) consists of terms describing the interaction of intralayer phononic fields with the electrons (the first term in Eq.(\ref{Equation_3})) via the coupling of the atomic displacement operators ${\rm{div{\hat{R}}}}_{\ell}\left({\bf{r}}\right)$ with the total electron densities in the layers $\hat{n}_{1}\left({\bf{r}}\right)$ and $\hat{n}_{2}\left({\bf{r}}\right)$. The parameter $\alpha$ before the sum represents the coupling constant between the electrons and phonons.     

The second term in Eq.(\ref{Equation_3}) represents the local Hubbard-$U$ interaction between electrons at a given lattice site. The third term in Eq.(\ref{Equation_3}) encapsulates the Coulomb interaction between electrons in different layers, characterized by the Coulomb interaction parameter $W$. The fourth term in Eq.(\ref{Equation_3}) describes the coupling between the electric field potentials $+V/2$ and $-V/2$ with the electron densities in the different layers. 
%
%
\begin{figure}
	\includegraphics[scale=0.3]{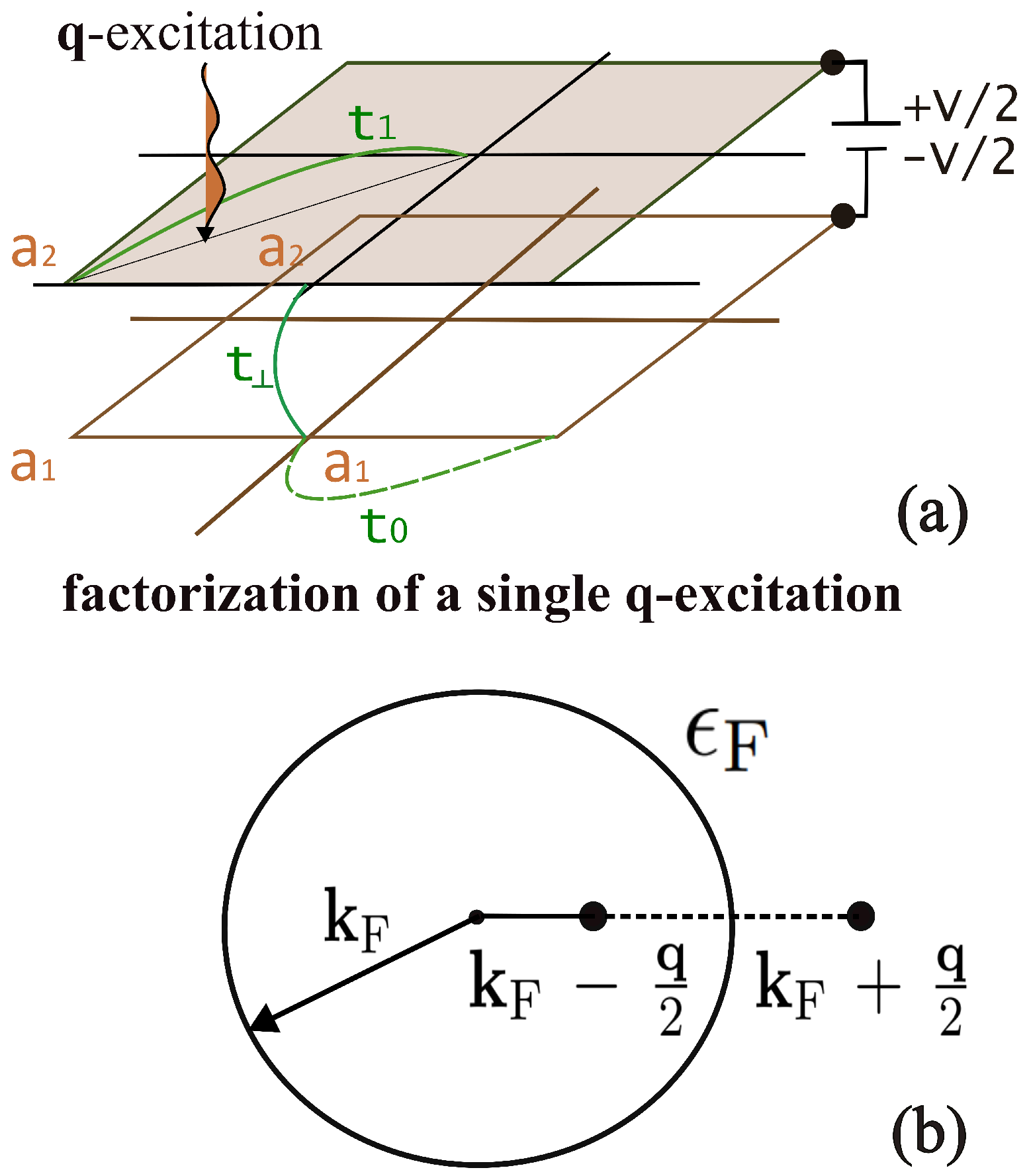}
	\caption{\label{fig:Fig_1}(Color online) \textbf{Panel (a):} Schematic representation of the metallic double-layer system with the applied electric field potential $V$. The upper layer is gated with an electric field potential of $V/2$, while the lower layer is gated with $-V/2$. The intralayer nearest-neighbor hopping amplitude ($t_0$), next-nearest neighbor hopping amplitude (nnn $t_1$), and interlayer hopping amplitude ($t_{\perp}$) are indicated in the picture. \textbf{Panel (b):} Factorization of a single-particle electron excitation using two simultaneous single-particle excitations, as described in Eq.(\ref{Equation_7}).}
\end{figure} 
%
%
The atomic displacement operators ${\hat{R}}_{\ell}\left({\bf{r}}\right)$ are associated with the phononic excitations and are defined as \cite{cite_26}
\begin{eqnarray}
{\hat{{\bf{R}}}}_{\ell}\left({\bf{r}}\tau\right)=\sum_{{\bf{q}}\Omega_{m}}\sqrt{\frac{\hbar}{2m\Omega_{{\bf{q}}}}}{\bf{e}}\left({\bf{q}}\right)\left[c_{\ell{\bf{q}}}\left(\Omega_{m}\right)e^{i\left({\bf{q}}{\bf{r}}-\Omega_{m}\tau\right)}\right.
\nonumber\\
\left.+c^{\dag}_{\ell{\bf{q}}}\left(\Omega_{m}\right)e^{-i\left({\bf{q}}{\bf{r}}-\Omega_{m}\tau\right)}\right],
\label{Equation_5}
\end{eqnarray} 
where $\hbar$ is the reduced Planck's constant, $m$ is the mass of the undressed phonon. The summation is over the phonon wave vectors ${\bf{q}}$ and the Bosonic Matsubara frequencies are $\Omega_{m}=2\pi{n}/\beta\hbar$, with $n$ being an integer and $\beta$ the inverse temperature given by $\beta=1/k_{\rm B}T$. Next, ${\bf{e}}\left({\bf{q}}\right)$ is the unit in the direction of the phonon's propagation wave vector ${\bf{q}}$ and $\Omega_{\bf{q}}$ is the phononic frequency associated with the phononic reciprocal mode ${\bf{q}}$. The operators $c^{\dag}_{\ell{\bf{q}}}\left(\Omega_{m}\right)$ and $c_{\ell{\bf{q}}}\left(\Omega_{m}\right)$ are the creation and annihilation operators of the phonons in the metallic layers $\ell=1$ and $\ell=2$, defined for each phononic mode $\left({\bf{q}},\Omega_{m}\right)$. The coupling constant $\alpha$, in Eq.(\ref{Equation_3}), is given by: 
\begin{eqnarray}
\alpha=ea^{2}C,
\label{Equation_6}
\end{eqnarray}   
with $C=ZeN_{\rm ion}/V$, where $Z$ is the atomic number, and $N_{\rm ion}$ is the number of atoms in the volume $V$.

Then, we employ the concept introduced in Ref.\cite{cite_24} to factorize a single-particle excitation with wave vector ${\bf{q}}$ into two simultaneous single-particle excitations with opposite wave vectors ${\bf{q}}/2$ and $-{\bf{q}}/2$. We have the relation for the energy required for single-particle excitation (per unit volume of the sample) as follows:
\begin{eqnarray}
\Delta{\epsilon}=\frac{1}{V}\left(\epsilon\left({\bf{k}}_{\rm F}+{\bf{q}}\right)-\epsilon\left({\bf{k}}_{F}\right)\right)=\rho\left(\epsilon_{F}\right)\epsilon^{\left({\bf{q}}\right)}_{1}\epsilon^{\left({\bf{q}}\right)}_{2},
\label{Equation_7}
\end{eqnarray}
%
where ${\bf{k}}_{\rm F}$ is the Fermi wave vector, the quasienergies $\epsilon^{\left({\bf{q}}\right)}_{1}$ and $\epsilon^{\left({\bf{q}}\right)}_{2}$
are equal and
\begin{eqnarray}
\epsilon^{\left({\bf{q}}\right)}_{1}=\epsilon^{\left({\bf{q}}\right)}_{2}=\epsilon^{\left({\bf{q}}\right)}=\frac{\pi{e^{4}}}{2}\sqrt{\frac{|{\bf{q}}|}{V}}{\rm {Ry}}^{-1},
\label{Equation_8}
\end{eqnarray}
where ${\rm {Ry}}=me^{4}/2\hbar^{2}=13.6$ eV. The function $\rho\left(\varepsilon_{F}\right)=mk_{\rm F}/\pi^{2}\hbar^{2}$ in Eq.(\ref{Equation_7}) represents the density of states at Fermi surface. 
 
Throughout the paper, we use the units in which $a_{0}\equiv 1$, $e\equiv 1$, and $V=1$. Thus, we obtain the two-quasiparticle ${\bf{q}}$-excitation energy (above or below the Fermi surface) as follows:
\begin{eqnarray} 
\epsilon^{\left({\bf{q}}_{0}\right)}=0.037  \ \ {\rm eV}, 
\label{Equation_9}
\end{eqnarray}  
where we set $|{\bf{q}}_{0}|=\pi/3$. Later-on, we will see that the solution of self-consistent equations in the problem could be obtained only at this value of the wave vector. 
For further convenience, we rewrite the intralayer and interlayer Coulomb interaction terms in Eq.(\ref{Equation_3}) in a different form. Specifically, for the intralayer interaction terms, we have:
\begin{eqnarray}
U\sum_{{\bf{r}}\ell}\hat{n}_{\ell\uparrow}\left({\bf{r}}\right)\hat{n}_{\ell\downarrow}\left({\bf{r}}\right)=\frac{U}{4}\sum_{{\bf{r}}\ell}\left(\hat{n}^{2}_{\ell}\left({\bf{r}}\right)-{\hat{p}}^{2}_{{\rm z}\ell}\left({\bf{r}}\right)\right),
\label{Equation_10}
\end{eqnarray} 
where we have introduced the electron spin polarization operator $\hat{p}_{{\rm z}\ell}=\hat{n}_{\ell\uparrow}-\hat{n}_{\ell\downarrow}$ for both layers $\ell=1$ and $\ell=2$. For the interlayer Coulomb interaction between the layers, we get
\begin{eqnarray}
&&\frac{W}{2}\sum_{{\bf{r}}\sigma\sigma'}\left(\hat{n}_{1\sigma}\left({\bf{r}}\right)\hat{n}_{2\sigma'}\left({\bf{r}}\right)+\hat{n}_{2\sigma}\left({\bf{r}}\right)\hat{n}_{1\sigma'}\left({\bf{r}}\right)\right)
\nonumber\\
&&=W\sum_{{\bf{r}}\sigma\sigma'}\left(\hat{n}_{1\sigma}\left({\bf{r}}\right)+\hat{n}_{2\sigma}\left({\bf{r}}\right)\right)
\nonumber\\
&&-\frac{W}{2}\sum_{{\bf{r}}\sigma\sigma'}\left(|\zeta_{\sigma'\sigma}\left({\bf{r}}\right)|^{2}+\hat{\zeta}_{\sigma'\sigma}\left({\bf{r}}\right){\hat{\zeta}}^{\dag}_{\sigma'\sigma}\left({\bf{r}}\right)\right).
\label{Equation_11}
\end{eqnarray}
Here, we define the on-site interlayer excitonic operators $\hat{\zeta}_{\sigma'\sigma}\left({\bf{r}}\right)$ and ${\hat{\zeta}}^{\dag}_{\sigma'\sigma}\left({\bf{r}}\right)$:
\begin{eqnarray}
\hat{\zeta}_{\sigma'\sigma}\left({\bf{r}}\right)=\hat{a}^{\dag}_{2\sigma'}\left({\bf{r}}\right)\hat{a}_{1\sigma}\left({\bf{r}}\right),
\nonumber\\
{\hat{\zeta}}^{\dag}_{\sigma'\sigma}\left({\bf{r}}\right)=\hat{a}^{\dag}_{1\sigma}\left({\bf{r}}\right)\hat{a}_{2\sigma'}\left({\bf{r}}\right).
\label{Equation_12}
\end{eqnarray}
The partition function of the system is given by:
\begin{eqnarray}
Z=\int\prod_{\ell=1,2}\left(\left[{\cal{D}}\hat{a}^{\dag}_{\ell}{\cal{D}}\hat{a}_{\ell}\right]\right)e^{-\beta{\hat{\cal{H}}}}.
\label{Equation_13}
\end{eqnarray}
Next, in the exponential of Eq.(\ref{Equation_13}), we perform Hubbard-Stratonovich decoupling \cite{cite_27} for the bilinear fermionic terms in Eqs.(\ref{Equation_10}) and (\ref{Equation_11}). Specifically, they are
\begin{eqnarray}
&&\exp\left(\int^{\beta}_{0}d\tau\sum_{{\bf{r}}}i\hat{n}_{\ell}\left({\bf{r}}\tau\right)\frac{U}{4}i\hat{n}_{\ell}\left({\bf{r}}\tau\right)\right)=
\nonumber\\
&&=\int\left[{\cal{D}}\hat{V}_{\ell}\right]e^{-\frac{1}{U}\sum_{{\bf{r}}}{\hat{V}}^{2}_{\ell}\left({\bf{r}}\tau\right)+\int^{\beta}_{0}d\tau\sum_{{\bf{r}}}i\hat{n}_{\ell}\left({\bf{r}}\tau\right){\hat{V}}_{\ell}\left({\bf{r}}\tau\right)},
\nonumber\\
\label{Equation_14}
\end{eqnarray}
where $\hat{V}_{\ell}\left({\bf{r}}\right)$ is the external decoupling source potential introduced for each lattice site ${\bf{r}}$.
The integral on the right-hand side of Eq.(\ref{Equation_14}) can be evaluated using saddle-point method, leading to the following contribution to the total Hamiltonian:
\begin{eqnarray}
{{\hat{\cal{H}}}}_{\rm U}=\frac{U}{2}\sum_{{\bf{r}}\ell}{\hat{n}}_{\ell}\left({\bf{r}}\right)\left\langle {\hat{n}}_{\ell}\right\rangle.
\label{Equation_15}
\end{eqnarray}
In turn, the decoupling of the polarization bilinear term in Eq.(\ref{Equation_10}) is straightforward:
\begin{eqnarray}
&&\exp\left(\frac{U}{4}\sum_{{\bf{r}}}{\hat{{\cal{P}}}}^{2}_{z\ell}\left({\bf{r}}\right)\right)=
\nonumber\\
&&=\int\left[{\cal{D}}{\hat{\phi}}_{\ell}\right]e^{-\frac{1}{U}\sum_{{\bf{r}}}{\hat{\phi}}^{2}_{\ell}\left({\bf{r}}\right)+\sum_{{\bf{r}}}{\hat{{\cal{P}}}}^{2}_{z\ell}\left({\bf{r}}\right){\hat{\phi}}_{\ell}\left({\bf{r}}\right)}.
\nonumber\\
\label{Equation_16}
\end{eqnarray}
We obtain the contribution to the total Hamiltonian from the saddle point evaluation of the integral on the right-hand side in Eq.(\ref{Equation_16}):
\begin{eqnarray}
{{\hat{\cal{H}}}}_{{\rm P}}=-\frac{U}{2}\sum_{{\bf{r}}\ell}{\hat{{\cal{P}}}}_{z\ell}\left({\bf{r}}\right)\left\langle {\hat{{\cal{P}}}}_{z\ell}\right\rangle.
\label{Equation_17}
\end{eqnarray}
Next, we introduce the fermionic complex Grassmann variables and express the partition function of the system in terms of these algebraic variables
\begin{eqnarray}
{\cal{Z}}=\int\prod_{\ell=1,2}\left(\left[{\cal{D}}\bar{a}_{\ell}{\cal{D}}a_{\ell}\right]\right)e^{-{\cal{S}}\left[\bar{a}_{1},a_{1}, \bar{a}_{2},a_{2}\right]/\hbar},
\label{Equation_18}
\end{eqnarray}
where ${\cal{S}}\left[\bar{a}_{1},a_{1}, \bar{a}_{2},a_{2}\right]$ in the exponential in Eq.(\ref{Equation_13}) is the total fermionic action of the system of two coupled metallic layers, and $\bar{a}_{\ell}\left({\bf{r}}\tau\right)$ and $a_{\ell}\left({\bf{r}}\tau\right)$ (with $\ell=1,2$) are fermionic Grassmann variables. In the Matsubara time path-integral formalism \cite{cite_27}, it is defined as:
\begin{eqnarray}
{\cal{S}}\left[\bar{a}_{1},a_{1}, \bar{a}_{2},a_{2}\right]=\int^{\beta}_{0}d\tau \sum_{{\bf{r}}}{\cal{H}}\left(\tau\right),
\label{Equation_19}
\end{eqnarray}
with the integration variable $\tau$, which is the imaginary time variable ($\tau\in\left[0,\beta\right]$), where $\beta$ is the inverse temperature given by $\beta=1/k_{\rm B}T$. 
After passing to Grassmann variables, the second term on the right-hand side of Eq.(\ref{Equation_11}) transforms into
\begin{eqnarray}
-\frac{W}{2}\sum_{{\bf{r}}\sigma\sigma'}\left(|\hat{\zeta}_{\sigma'\sigma}\left({\bf{r}}\right)|^{2}+\hat{\zeta}_{\sigma'\sigma}\left({\bf{r}}\right){\hat{\zeta}}^{\dag}_{\sigma'\sigma}\left({\bf{r}}\right)\right)\rightarrow
\nonumber\\
\rightarrow-W\sum_{{\bf{r}}\sigma\sigma'}|\zeta_{\sigma'\sigma}\left({\bf{r}}\right)|^{2},
\label{Equation_20}
\end{eqnarray}
where
\begin{eqnarray}
|\zeta_{\sigma'\sigma}\left({\bf{r}}\right)|^{2}={\bar{\zeta}}_{\sigma'\sigma}\left({\bf{r}}\right){\zeta}_{\sigma'\sigma}\left({\bf{r}}\right)
\label{Equation_21}
\end{eqnarray}
and $\bar{\zeta}_{\sigma'\sigma}\left({\bf{r}}\right)$, $\zeta_{\sigma'\sigma}\left({\bf{r}}\right)$ are complex Grassmann variables defined analogously to the operators in Eq.(\ref{Equation_12}):
\begin{eqnarray}
{\zeta}_{\sigma'\sigma}\left({\bf{r}}\right)=\bar{a}_{2\sigma'}\left({\bf{r}}\right){a}_{1\sigma}\left({\bf{r}}\right),
\nonumber\\
{\bar{\zeta}}_{\sigma'\sigma}\left({\bf{r}}\right)=\bar{a}_{1\sigma}\left({\bf{r}}\right)a_{2\sigma'}\left({\bf{r}}\right).
\label{Equation_22}
\end{eqnarray}
The Hubbard-Stratonovich transformation for the bilinear form in Eq.(\ref{Equation_20}) is: 
\begin{eqnarray}
\exp\left(W\int^{\beta}_{0}d\tau\sum_{{\bf{r}}}|\zeta_{\sigma'\sigma}\left({\bf{r}}\tau\right)|^{2}\right)=
\nonumber\\
=\int\left[{\cal{D}}\bar{\Xi}{\cal{D}}\Xi\right]e^{-\frac{1}{W}\int^{\beta}_{0}d\tau\sum_{{\bf{r}}\sigma\sigma'}|\Xi_{\sigma'\sigma}\left({\bf{r}}\tau\right)|^{2}}\times
\nonumber\\
\times e^{\int^{\beta}_{0}d\tau\sum_{{\bf{r}}}\bar{\zeta}_{\sigma'\sigma}\left({\bf{r}}\tau\right)\Xi_{\sigma'\sigma}\left({\bf{r}}\tau\right)+\bar{\Xi}_{\sigma'\sigma}\left({\bf{r}}\tau\right){\zeta}_{\sigma'\sigma}\left({\bf{r}}\tau\right)}.
\label{Equation_23}
\end{eqnarray}
After the saddle-point evaluation of the integral on the right-hand side of Eq.(\ref{Equation_23}), we obtain the mean-field values of the decoupling fields $\bar{\zeta}_{\sigma'\sigma}\left({\bf{r}}\tau\right)$ and ${\zeta}_{\sigma'\sigma}\left({\bf{r}}\tau\right)$:
\begin{eqnarray}
\bar{\zeta}_{\rm s}=W\left\langle \bar{a}_{1\sigma}\left({\bf{r}}\tau\right)a_{2\sigma'}\left({\bf{r}}\tau\right)\right\rangle,
\nonumber\\
{\zeta}_{\rm s}=W\left\langle \bar{a}_{2\sigma'}\left({\bf{r}}\tau\right){a}_{1\sigma}\left({\bf{r}}\tau\right)\right\rangle.
\label{Equation_24}
\end{eqnarray}
The obtained values are, in fact, the excitonic order parameters in the system, denoted as $\bar{\Delta}_{\sigma'\sigma}$ and ${\Delta}_{\sigma'\sigma}$. The contribution to the Hamiltonian arising from these terms is:
\begin{eqnarray}
{\cal{H}}_{\rm W}=-\sum_{{\bf{r}}\sigma}\left(\bar{\Delta}_{\sigma}\bar{a}_{2\sigma}\left({\bf{r}}\right){a}_{1\sigma}\left({\bf{r}}\right)+{\Delta}_{\sigma}\bar{a}_{1\sigma}\left({\bf{r}}\right)a_{2\sigma}\left({\bf{r}}\right)\right).
\nonumber\\
\label{Equation_25}
\end{eqnarray}
We suppose here, that there is no antiferromagnetic order in the layers, such that $\left\langle {\hat{{\cal{P}}}}_{z\ell}\right\rangle=0$ for $\ell=1,2$. 
Then, the total partition function of the system becomes:
\begin{eqnarray}
{\cal{S}}={\cal{S}}_{\rm 0}+{\cal{S}}_{\rm U}+{\cal{S}}_{\rm W}+{\cal{S}}_{\rm g}+{\cal{S}}_{\rm int},
\label{Equation_26}
\end{eqnarray}
where
\begin{widetext}
\begin{eqnarray}
&&{\cal{S}}_{\rm 0}=\int^{\beta}_{0}d\tau {\cal{H}}_{\rm 0}\left(\tau\right),
\nonumber\\
&&{\cal{S}}_{\rm U}=\int^{\beta}_{0}d\tau {\cal{H}}_{\rm U}\left(\tau\right)=\frac{U}{2}\sum_{{\bf{r}}\ell}\int^{\beta}_{0}d\tau{\hat{n}}_{\ell}\left({\bf{r}}\tau\right)\left\langle {\hat{n}}_{\ell}\right\rangle,
\nonumber\\
&&{\cal{S}}_{\rm W}=\int^{\beta}_{0}d\tau {\cal{H}}_{\rm W}\left(\tau\right)=-\sum_{{\bf{r}}\sigma}\int^{\beta}_{0}d\tau\left(\bar{\Delta}_{\sigma}\bar{a}_{2\sigma}\left({\bf{r}}\tau\right){a}_{1\sigma}\left({\bf{r}}\tau\right)+{\Delta}_{\sigma}\bar{a}_{1\sigma}\left({\bf{r}}\tau\right)a_{2\sigma}\left({\bf{r}}\tau\right)\right),
\nonumber\\
&&{\cal{S}}_{\rm g}=-\alpha\int^{\beta}_{0}d\tau\sum_{{\bf{r}}\ell}{n}_{\ell}\left({\bf{r}}\tau\right){\rm{div{R}}}_{\ell}\left({\bf{r}}\tau\right),
\nonumber\\
&&{\cal{S}}_{\rm int}=\frac{V}{2}\int^{\beta}_{0}d\tau\sum_{{\bf{r}}\ell}\left({n}_{2}\left({\bf{r}}\tau\right)-{n}_{1}\left({\bf{r}}\tau\right)\right).
\label{Equation_27}
\end{eqnarray}
\end{widetext}
%
\subsection{\label{sec:Section_2_2} The effective electron-phonon interaction}
%
In order to retrieve the effective electron-phonon interaction in the system of two metallic layers, we consider the last free phonon term in the free Hamiltonian in Eq.(\ref{Equation_2}), which appears also in the action ${\cal{S}}_{0}$ in Eq.(\ref{Equation_26}), the action ${\cal{S}}_{\rm el-ph}$, and the free phonon's Berry term. We then express the corresponding electron-phonon interaction action in Fourier space:
%
%
\begin{figure}
	\includegraphics[scale=0.6]{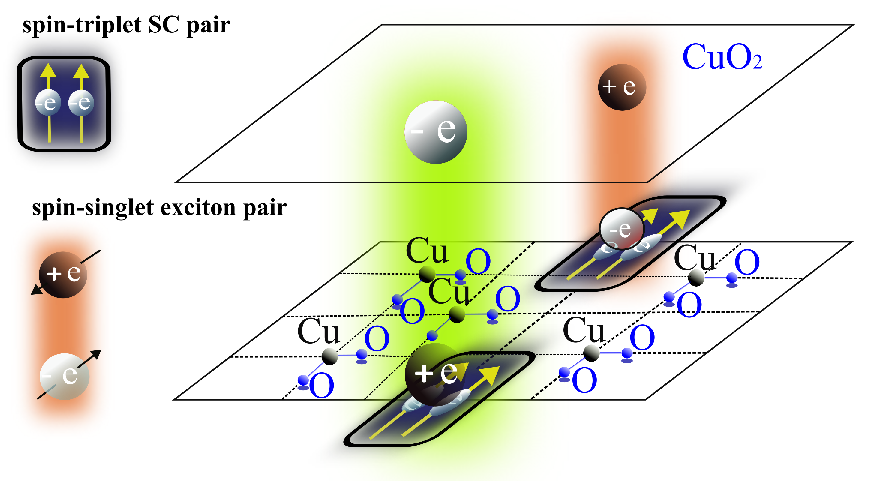}
	\caption{\label{fig:Fig_2}(Color online) As a complement to Fig.~\ref{fig:Fig_1}, we illustrate the metallic double-layer system with excitonic pair formations (depicted in yellow and red) and superconducting pair formations (represented by the opaque squares on the top and bottom layers). On the left side of the figure, the spin-triplet superconducting pairs and spin-singlet excitonic pairs are shown. }
\end{figure} 
%
%
\begin{eqnarray}
{\cal{S}}_{\rm{el-ph}}=-\beta{\hbar}\sum_{{\bf{q}}\Omega_{m}\ell}\hbar\Omega_{{\bf{q}}}\bar{c}_{\ell{\bf{q}}}\left(\Omega_{m}\right){c}_{\ell{\bf{q}}}\left(\Omega_{m}\right)
\nonumber\\
+\beta\hbar\sum_{{\bf{q}}\Omega_{m}\ell}\bar{c}_{\ell{\bf{q}}}\left(\Omega_{m}\right)\left(-i\Omega_{m}\hbar\right){c}_{\ell{\bf{q}}}\left(\Omega_{m}\right)
\nonumber\\
-\hbar\sum_{{\bf{q}}\Omega_{m}\ell}\left(\bar{\gamma}_{\ell{\bf{q}}}\left(\Omega_{m}\right){c}_{\ell{\bf{q}}}\left(\Omega_{m}\right)+\bar{c}_{\ell{\bf{q}}}\left(\Omega_{m}\right){\gamma}_{\ell{\bf{q}}}\left(\Omega_{m}\right)\right),
\label{Equation_28}
\end{eqnarray}
where $\Omega_{{\bf{q}}}$ is the single phonon frequency with the wave vector ${\bf{q}}$, $\Omega_{m}$ are bosoninc Matsubara frequencies $\Omega_{m}=2\pi{m}/{\beta\hbar}$ with $m$ being the integer numbers $m=0, \pm1, \pm2, ...$. The functions $\bar{\gamma}_{\ell{\bf{q}}}\left(\Omega_{m}\right)$ and ${\gamma}_{\ell{\bf{q}}}\left(\Omega_{m}\right)$ in the third electron-phonon interaction terms are 
\begin{widetext}
\begin{eqnarray}
\bar{\gamma}_{\ell{\bf{q}}}\left(\Omega_{m}\right)=ig_{{\bf{q}}}\sum_{{\bf{k}}\nu_{n},\sigma}\left[\bar{a}_{\ell\sigma}\left({\bf{k}}+\frac{{\bf{q}}}{2},\nu_{n}+\frac{\Omega_{m}}{2}\right){a}_{\ell\sigma}\left({\bf{k}}-\frac{{\bf{q}}}{2},\nu_{n}-\frac{\Omega_{m}}{2}\right)\right],
\nonumber\\
{\gamma}_{\ell{\bf{q}}}\left(\Omega_{m}\right)=-ig_{{\bf{q}}}\sum_{{\bf{k}}\nu_{n},\sigma}\left[\bar{a}_{\ell\sigma}\left({\bf{k}}-\frac{{\bf{q}}}{2},\nu_{n}-\frac{\Omega_{m}}{2}\right){a}_{\ell\sigma}\left({\bf{k}}+\frac{{\bf{q}}}{2},\nu_{n}+\frac{\Omega_{m}}{2}\right)\right].
\nonumber\\
\label{Equation_29}
\end{eqnarray}
\end{widetext}
The dimensionless coupling constant $g_{{\bf{q}}}$ is defined as 
\begin{eqnarray}
g_{{\bf{q}}}=\alpha\beta\sqrt{\frac{\hbar}{2m\Omega_{{\bf{q}}}}}|{\bf{q}}|.
\label{Equation_30}
\end{eqnarray}
In the case of non-dressed phonons, the coupling constant in Eq.(\ref{Equation_28}) simplifies to $g_{{\bf{q}}}=\alpha\beta$.
Next, we integrate out the phonon field using the Hubbard-Stratonovich transformation for the phonons:
\begin{widetext}
\begin{eqnarray}
\int\left[{\cal{D}}\bar{c}_{\ell}{\cal{D}}c_{\ell}\right]e^{-\beta\sum_{{\bf{q}}\Omega_{m}}\bar{c}_{\ell{\bf{q}}}\left(\Omega_{m}\right)\left(-\hbar\Omega_{{\bf{q}}}-i\hbar\Omega_{m}\right)c_{\ell{\bf{q}}}\left(\Omega_{m}\right)+\sum_{{\bf{q}}\Omega_{m}}\left(\bar{\gamma}_{\ell\bf{q}}\left(\Omega_{m}\right)c_{\ell\bf{q}}\left(\Omega_{m}\right)+\bar{c}_{\ell\bf{q}}\left(\Omega_{m}\right)\gamma_{\ell\bf{q}}\left(\Omega_{m}\right)\right)}=e^{-\sum_{{\bf{q}}\Omega_{m}}\frac{\left(\beta\hbar\Omega_{{\bf{q}}}-i\beta\hbar\Omega_{m}\right)|\gamma_{\ell\bf{q}}\left(\Omega_{m}\right)|^{2}}{\left[\left(\beta\hbar\Omega_{q}\right)^{2}+\left(\beta\hbar\Omega_{m}\right)^{2}\right]}}.
\label{Equation_31}
\end{eqnarray}
\end{widetext}
In the limit of small phononic excitations, i.e., when $\hbar\Omega_{\bf{q}}\rightarrow 0$, we obtain the Lorentzian form in the exponential on the right-hand side of Eq.(\ref{Equation_29}), and only the mode $\Omega_{m}=0$ survives in the sum over the frequencies $\Omega_{m}$:
\begin{eqnarray}  
\frac{\beta\hbar\Omega_{\bf{q}}}{\left(\beta\hbar\Omega_{q}\right)^{2}+\left(\beta\hbar\Omega_{m}\right)^{2}}=\pi\delta\left(\beta\hbar\Omega_{m}\right)
\label{Equation_32}
\end{eqnarray}
and we have 
\begin{eqnarray}
\sum_{{\bf{q}}\Omega_{m}}\frac{\beta\hbar\Omega_{\bf{q}}|\gamma_{\bf{q}}\left(\Omega_{m}\right)|^{2}}{\left(\beta\hbar\Omega_{q}\right)^{2}+\left(\beta\hbar\Omega_{m}\right)^{2}}=\pi\sum_{{\bf{q}}}|\gamma_{\bf{q}}\left(\Omega_{m}=0\right)|^{2}.
\label{Equation_33}
\end{eqnarray}
The second sum is:
\begin{eqnarray}
-\sum_{{\bf{q}}\Omega_{m}}\frac{\beta\hbar\Omega_{m}|\gamma_{\bf{q}}\left(\Omega_{m}\right)|^{2}}{\left(\beta\hbar\Omega_{q}\right)^{2}+\left(\beta\hbar\Omega_{m}\right)^{2}}=0
\label{Equation_34}
\end{eqnarray}
due to the parity of the function $\frac{|\gamma_{\bf{q}}\left(\Omega_{m}\right)|^{2}}{\left(\beta\hbar\Omega_{q}\right)^{2}+\left(\beta\hbar\Omega_{m}\right)^{2}}$ when multiplied with $\Omega_{m}$. Then, the action in Eq.(\ref{Equation_28}) transforms into:
\begin{eqnarray}
{\cal{S}}_{\rm{el-ph}}=-\hbar\pi\alpha^{2}\beta^{2}\sum_{\substack{{\bf{k}}{\bf{q}}\\ \nu_{n}\sigma}}\left(|\Lambda^{\left(1\right)}_{{\bf{k}}{\bf{q}}\sigma}\left(\nu_{n}\right)|^{2}+|\Lambda^{\left(2\right)}_{{\bf{k}}{\bf{q}}\sigma}\left(\nu_{n}\right)|^{2}\right),
\nonumber\\
\label{Equation_35}
\end{eqnarray}
where $\bar{\Lambda}^{\left(\ell\right)}_{{\bf{k}}{\bf{q}}\sigma}\left(\nu_{n}\right)$ and $\Lambda^{\left(\ell\right)}_{{\bf{k}}{\bf{q}}\sigma}\left(\nu_{n}\right)$ (with $\ell=1,2$) are defined as follows:
\begin{eqnarray}
\Lambda^{\left(\ell\right)}_{{\bf{k}}{\bf{q}}\sigma}\left(\nu_{n}\right)=a_{\ell\sigma}\left({\bf{k}}-\frac{{\bf{q}}}{2},\nu_{n}\right)a_{\ell\sigma}\left({\bf{k}}+\frac{{\bf{q}}}{2}\right),
\nonumber\\
\bar{\Lambda}^{\left(\ell\right)}_{{\bf{k}}{\bf{q}}\sigma}\left(\nu_{n}\right)=\bar{a}_{\ell\sigma}\left({\bf{k}}+\frac{{\bf{q}}}{2},\nu_{n}\right)\bar{a}_{\ell\sigma}\left({\bf{k}}-\frac{{\bf{q}}}{2}\right).
\label{Equation_36}
\end{eqnarray}
Furthermore, we decouple the biquadratic fermionic terms in Eq.(\ref{Equation_35}) using an algebraic Hubbard-Stratonovich transformation:
\begin{widetext}
\begin{eqnarray}
\exp\left(\pi\alpha^{2}\beta^{2}\sum_{\substack{{\bf{k}}{\bf{q}}\\ \nu_{n}\sigma}}|\Lambda^{\left(\ell\right)}_{{\bf{k}}{\bf{q}}\sigma}\left(\nu_{n}\right)|^{2}\right)=\int\left[{{\cal{D}}\bar{\Psi}^{\left(\ell\right)}_{\rm s}{\cal{D}}\Psi^{\left(\ell\right)}_{\rm s}}\right]e^{-\frac{1}{\pi\alpha^{2}\beta^{2}}\sum_{\substack{{\bf{k}}{\bf{q}}\\ \nu_{n}\sigma}}|\Psi^{\left(\ell\right)}_{s{\bf{k}}{\bf{q}}\sigma}\left(\nu_{n}\right)|^{2}+\sum_{\substack{{\bf{k}}{\bf{q}}\\ \nu_{n}\sigma}}\left[\bar{\Psi}^{\left(\ell\right)}_{s{{\bf{k}}{\bf{q}}\sigma}}\left(\nu_{n}\right)\Lambda^{\left(\ell\right)}_{{\bf{k}}{\bf{q}}\sigma}\left(\nu_{n}\right)+\bar{\Lambda}^{\left(\ell\right)}_{{\bf{k}}{\bf{q}}\sigma}\left(\nu_{n}\right){\Psi}^{\left(\ell\right)}_{s{{\bf{k}}{\bf{q}}\sigma}}\left(\nu_{n}\right)\right]}.
\nonumber\\
\label{Equation_37}
\end{eqnarray}
\end{widetext}
The saddle-point evaluation of the exponential on the right-hand side of Eq.(\ref{Equation_37}) yields the expectation values for the decoupling fields $\bar{\Psi}^{\left(\ell\right)}_{{\rm s}{{\bf{k}}{\bf{q}}\sigma}}\left(\nu_{n}\right)$ and ${\Psi}^{\left(\ell\right)}_{{\rm s}{{\bf{k}}{\bf{q}}\sigma}}\left(\nu_{n}\right)$. We have:
\begin{eqnarray}
{\Psi}^{\left(\ell\right)}_{\rm{s,m}}=\pi\alpha^{2}\beta^{2}\left\langle \Lambda^{\left(\ell\right)}_{{\bf{k}}{\bf{q}}\sigma}\left(\nu_{n}\right)\right\rangle,
\nonumber\\
\bar{\Psi}^{\left(\ell\right)}_{\rm{s,m}}=\pi\alpha^{2}\beta^{2}\left\langle \bar{\Lambda}^{\left(\ell\right)}_{{\bf{k}}{\bf{q}}\sigma}\left(\nu_{n}\right)\right\rangle.
\label{Equation_38}
\end{eqnarray}
At this point, we define the superconducting order parameters in both layers as $\bar{\Delta}^{\ell}_{{\rm s}}$ and ${\Delta}^{\ell}_{{\rm s}}$
\begin{eqnarray}
\bar{\Delta}^{\ell}_{{\rm s}}=\pi\beta\alpha^{2}\left\langle \bar{\Lambda}^{\left(\ell\right)}_{{\bf{k}}{\bf{q}}\sigma}\left(\nu_{n}\right)\right\rangle,
\nonumber\\
{\Delta}^{\ell}_{{\rm s}}=\pi\beta\alpha^{2}\left\langle \Lambda^{\left(\ell\right)}_{{\bf{k}}{\bf{q}}\sigma}\left(\nu_{n}\right)\right\rangle.
\label{Equation_39}
\end{eqnarray}
%
\subsection{\label{sec:Section_2_3} Total action and self-consistent equations}
%
The contribution to the total action of the system, arising from the integration over phonons, is
\begin{eqnarray}
{\cal{S}}_{\rm el-ph}=&&-\beta\hbar\sum_{\substack{{\bf{k}}{\bf{q}}\\ \nu_{n}\sigma}}\sum_{\ell}\left[\bar{\Delta}^{\ell}_{{\rm s}}a_{\ell\sigma}\left({\bf{k}}-\frac{{\bf{q}}}{2},\nu_{n}\right)a_{\ell\sigma}\left({\bf{k}}+\frac{{\bf{q}}}{2}\right)\right.
\nonumber\\
&&\left.+\bar{a}_{\ell\sigma}\left({\bf{k}}+\frac{{\bf{q}}}{2},\nu_{n}\right)\bar{a}_{\ell\sigma}\left({\bf{k}}-\frac{{\bf{q}}}{2}\right){\Delta}^{\ell}_{{\rm s}}\right].
\label{Equation_40}
\end{eqnarray}
Here, we make the principal assumption in the problem concerning the average charge densities in the layers, denoted as $\bar{n}_{\ell}$ (where the statistical average $\left\langle\hat{n}_{\ell}\right\rangle$ is replaced by $\bar{n}_{\ell}$ for the Grassmann variables). Specifically, we introduce doping into the system as follows: 
\begin{eqnarray}
\bar{n}_{a1}+\bar{n}_{a2}=\frac{1}{\kappa}.
\label{Equation_41}
\end{eqnarray}
The value $\kappa = 0.5$ corresponds to half-filling, representing the situation where the maximum allowable number of electrons at each atomic site is $1$. Therefore, $\kappa_{\text{min}} = 0.25$. Values $\kappa < \kappa_{\text{min}}$ are excluded due to the violation of the Pauli exclusion principle. Conversely, when $\kappa > 0.25$, it indicates a doped bilayer case, and we define the doping in the system as:
\begin{eqnarray}
x=\frac{1}{\kappa_{\rm min}}-\frac{1}{\kappa}. 
\label{Equation_42}
\end{eqnarray}
For instance, if $\kappa \in \left[0.25,1.0\right]$, the electron doping $x$ would then lie in the range $x\in\left[0, 1.0\right]$. We also introduce the average charge imbalance between the metallic layers as:
\begin{eqnarray}
\bar{n}_{a2}-\bar{n}_{a1}=\delta{\bar{n}}.
\label{Equation_43}
\end{eqnarray}
Thus, from Eqs.(\ref{Equation_41}) and (\ref{Equation_43}), we obtain the average electron charge densities in terms of $\kappa$ and $\delta{\bar{n}}$:
\begin{eqnarray}
\bar{n}_{a1}=0.5\left({\kappa}^{-1}-\delta{\bar{n}}\right),
\nonumber\\
\bar{n}_{a2}=0.5\left({\kappa}^{-1}+\delta{\bar{n}}\right).
\label{Equation_44}
\end{eqnarray}

Furthermore, we express the other terms in Eq.(\ref{Equation_27}) in Fourier representation and we introduce the following Nambu spinors for the density coupling mechanism:
\begin{eqnarray} 
{\Psi}_{{\bf{k}}{\bf{q}}\sigma}\left(\nu_{n}\right)=\left(
\begin{array}{crrrr}
\bar{a}_{1\sigma}\left({\bf{k}}-\frac{{\bf{q}}}{2},\nu_{n}\right)\\\\
{a}_{1\sigma}\left({\bf{k}}+\frac{{\bf{q}}}{2},\nu_{n}\right) \\\\
\bar{a}_{2\sigma}\left({\bf{k}}-\frac{{\bf{q}}}{2},\nu_{n}\right) \\\\
a_{2\sigma}\left({\bf{k}}+\frac{{\bf{q}}}{2},\nu_{n}\right) \\\\
\end{array}\right).
\label{Equation_45}
\end{eqnarray}
The corresponding complex conjugate spinor is denoted as: 
\begin{widetext}
\begin{eqnarray} 
\bar{\Psi}_{{\bf{k}}{\bf{q}}\sigma}\left(\nu_{n}\right)=\left({a}_{1\sigma}\left({\bf{k}}-\frac{{\bf{q}}}{2},\nu_{n}\right),\bar{a}_{1\sigma}\left({\bf{k}}+\frac{{\bf{q}}}{2},\nu_{n}\right), {a}_{2\sigma}\left({\bf{k}}-\frac{{\bf{q}}}{2},\nu_{n}\right),\bar{a}_{2\sigma}\left({\bf{k}}+\frac{{\bf{q}}}{2},\nu_{n}\right)\right).
\label{Equation_46}
\end{eqnarray}
\end{widetext}
Then, the action of the system can be expressed using the Gorkov matrix as follows:
\begin{eqnarray} 
{\cal{S}}=\frac{\beta\hbar}{2}\sum_{{\bf{k}}{\bf{q}}}\sum_{\sigma\nu_{n}}\bar{\Psi}_{{\bf{k}}{\bf{q}}\sigma}\left(\nu_{n}\right){\cal{G}}^{-1}_{{\bf{k}}{\bf{q}}\sigma}\left(\nu_{n}\right){\Psi}_{{\bf{k}}{\bf{q}}\sigma}\left(\nu_{n}\right),
\label{Equation_47}
\end{eqnarray}
where ${\cal{G}}^{-1}_{{\bf{k}}{\bf{q}}\sigma}\left(\nu_{n}\right)$ is the Gorkov matrix for the considered problem
\begin{eqnarray}
\noindent 
	\resizebox{\linewidth}{!}{\columnsep=1pt %
	${\cal{G}}^{-1}_{{\bf{k}}{\bf{q}}\sigma}\left(\nu_{n}\right)=\left(
     \begin{matrix}
		-i\nu_{n}+\mu_{1{\bf{k}}{\bf{q}}}& -2\bar{\Delta}_{\rm{s}\sigma} & \bar{\Delta}_{\sigma}+t_\perp & 0\\
		-2{\Delta}_{\rm{s}\sigma} &i\nu_{n}-\mu_{2{\bf{k}}{\bf{q}}}&  0 & -{\Delta}_{\sigma}-t_\perp\\
		{\Delta}_{\sigma}+t_\perp & 0 & -i\nu_{n}+\mu_{3{\bf{k}}{\bf{q}}} & -2\bar{\Delta}_{\rm{s}\sigma} \\
		0 & -\bar{\Delta}_{\sigma}-t_\perp & -2{\Delta}_{\rm{s}\sigma} & i\nu_{n}-\mu_{4{\bf{k}}{\bf{q}}}\\
	\end{matrix}\right)$}.
	\nonumber\\
	\label{Equation_48}
\end{eqnarray}
We have introduced the effective chemical potentials $\mu_{1}$ and $\mu_{2}$ in Eq.(\ref{Equation_48}) as follows:
\begin{eqnarray}
\mu_{1{\bf{k}}{\bf{q}}}=\mu_{0}+2\tilde{\gamma}\left({\bf{k}}-\frac{\bf{q}}{2}\right)+\frac{U}{4}\delta{\bar{n}}+\frac{V}{2}+W,
\nonumber\\
\mu_{2{\bf{k}}{\bf{q}}}=\mu_{0}+2\tilde{\gamma}\left({\bf{k}}+\frac{\bf{q}}{2}\right)+\frac{U}{4}\delta{\bar{n}}+\frac{V}{2}+W,
\nonumber\\
\mu_{3{\bf{k}}{\bf{q}}}=\mu_{0}+2\tilde{\gamma}\left({\bf{k}}-\frac{\bf{q}}{2}\right)-\frac{U}{4}\delta{\bar{n}}-\frac{V}{2}+W,
\nonumber\\
\mu_{4{\bf{k}}{\bf{q}}}=\mu_{0}+2\tilde{\gamma}\left({\bf{k}}+\frac{\bf{q}}{2}\right)-\frac{U}{4}\delta{\bar{n}}-\frac{V}{2}+W,
\label{Equation_49}
\end{eqnarray}
where 
\begin{eqnarray}
&&\mu_{0}=\mu-\frac{U}{4\kappa},
\nonumber\\
&&\tilde{\gamma}\left({\bf{k}}-\frac{\bf{q}}{2}\right)=2t_{0}\left(\cos\left({\bf{k}}-\frac{\bf{q}}{2}\right)+\cos\left({\bf{k}}-\frac{\bf{q}}{2}\right)\right)
\nonumber\\
&&+4t_{1}\cos\left({\bf{k}}-\frac{\bf{q}}{2}\right)\cos\left({\bf{k}}-\frac{\bf{q}}{2}\right),
\nonumber\\
&&\tilde{\gamma}\left({\bf{k}}+\frac{\bf{q}}{2}\right)=2t_{0}\left(\cos\left({\bf{k}}+\frac{\bf{q}}{2}\right)+\cos\left({\bf{k}}+\frac{\bf{q}}{2}\right)\right)
\nonumber\\
&&+4t_{1}\cos\left({\bf{k}}+\frac{\bf{q}}{2}\right)\cos\left({\bf{k}}+\frac{\bf{q}}{2}\right).
\nonumber\\
\label{Equation_50}
\end{eqnarray}

We solve the equation $\det{{\cal{G}}^{-1}_{{\bf{k}}{\bf{q}}\sigma}\left(\nu_{n}\right)}=0$ and we find the electronic band structure energies:
\begin{eqnarray}
\epsilon_{1{\bf{k}}{\bf{q}}}=0.5\left(-2W+\tilde{\mu}_1+\tilde{\mu}_2-\sqrt{\alpha_{{\bf{k}}{\bf{q}}}-4\sqrt{\xi_{{\bf{k}}{\bf{q}}}}}\right),
\nonumber\\
\epsilon_{2{\bf{k}}{\bf{q}}}=0.5\left(-2W+\tilde{\mu}_1+\tilde{\mu}_2+\sqrt{\alpha_{{\bf{k}}{\bf{q}}}-4\sqrt{\xi_{{\bf{k}}{\bf{q}}}}}\right),
\nonumber\\
\epsilon_{3{\bf{k}}{\bf{q}}}=0.5\left(-2W+\tilde{\mu}_1+\tilde{\mu}_2-\sqrt{\alpha_{{\bf{k}}{\bf{q}}}+4\sqrt{\xi_{{\bf{k}}{\bf{q}}}}}\right),
\nonumber\\
\epsilon_{4{\bf{k}}{\bf{q}}}=0.5\left(-2W+\tilde{\mu}_1+\tilde{\mu}_2+\sqrt{\alpha_{{\bf{k}}{\bf{q}}}+4\sqrt{\xi_{{\bf{k}}{\bf{q}}}}}\right).
\nonumber\\
\label{Equation_51}
\end{eqnarray}
These eigenvalues are the same for both spin directions, $\sigma=\uparrow$ and $\sigma=\downarrow$, due to the absence of antiferromagnetic order in the system. We have:  
\begin{eqnarray}
&&\tilde{\mu}_1=\mu_{0}+2\tilde{\gamma}\left({\bf{k}}-\frac{\bf{q}}{2}\right)+a\left(U,V\right),
\nonumber\\
&&\tilde{\mu}_2=\mu_{0}+2\tilde{\gamma}\left({\bf{k}}+\frac{\bf{q}}{2}\right)+a\left(U,V\right),
\nonumber\\
&&\alpha_{{\bf{k}}{\bf{q}}}=4a^{2}\left(U,V\right)+4|\Delta_{\sigma}+t_0|^{2}-16|\Delta_{\rm{s}\sigma}|^{2}
\nonumber\\
&&+\left(\tilde{\mu}_1-\tilde{\mu}_2\right)^{2},
\nonumber\\
&&\xi_{{\bf{k}}{\bf{q}}}=-4\left[4a^{2}\left(U,V\right)+4|\Delta_{\sigma}+t_0|^{2}\right]|\Delta_{\rm{s}\sigma}|^{2}
\nonumber\\
&&+\left[a^{2}\left(U,V\right)+|\Delta_{\sigma}+t_0|^{2}\right]\left(\tilde{\mu}_1-\tilde{\mu}_{2}\right)^{2}
\nonumber\\
\label{Equation_52}
\end{eqnarray}
and
\begin{eqnarray}
a\left(U,V\right)=\frac{U}{4}\delta{\bar{n}}+\frac{V}{2}.
\label{Equation_53}
\end{eqnarray}
Next, we write the system of self-consistent equations for the chemical potential, average charge density difference between the layers, the excitonic order parameter, and the triplet superconducting order parameter. For the first two physical quantities, we have the equations in Eqs. (\ref{Equation_41}) and (\ref{Equation_43}), as presented in Section \ref{sec:Section_2_2}. Furthermore, the equations for the excitonic and superconducting order parameters are defined in Eqs.(\ref{Equation_24}) and (\ref{Equation_39}), and we assume that the excitonic and superconducting order parameters are real. The system of equations in Fourier space will have the form:
\begin{eqnarray}
\frac{2}{N_{\rm ph}}\sum_{{\bf{q}}}\sum^{4}_{i=1}\alpha_{i{\bf{k}}_{0}{\bf{q}}}n_{\rm F}\left(\mu-\epsilon_{i{\bf{k}}_{0}{\bf{q}}}\right)=\frac{1}{\kappa},
\nonumber\\
\frac{2}{N_{\rm ph}}\sum_{{\bf{q}}}\sum^{4}_{i=1}\beta_{i{\bf{k}}_{0}{\bf{q}}}n_{\rm F}\left(\mu-\epsilon_{i{\bf{k}}_{0}{\bf{q}}}\right)=\delta{\bar{n}},
\nonumber\\
\frac{W\left(\Delta_{\sigma}+t_{\perp}\right)}{N_{\rm ph}}\sum_{{\bf{q}}}\sum^{4}_{i=1}\gamma_{i{\bf{k}}_{0}{\bf{q}}}n_{\rm F}\left(\mu-\epsilon_{i{\bf{k}}_{0}{\bf{q}}}\right)=\Delta_{\sigma},
\nonumber\\
\frac{g\Delta_{\rm s\sigma}}{N_{\rm ph}}\sum_{{\bf{q}}}\sum^{4}_{i=1}\delta_{i{\bf{k}}_{0}{\bf{q}}}n_{\rm F}\left(\mu-\epsilon_{i{\bf{k}}_{0}{\bf{q}}}\right)=\Delta_{\rm s\sigma},
\label{Equation_54}
\end{eqnarray}
where $N_{\rm ph}$ is the number of wave vectors ${\bf{q}}$, $n_{F}\left(x\right)$ is the Fermi-Dirac distribution function
\begin{eqnarray}
n_{F}\left(x\right)=\frac{1}{e^{x-\mu}+1}
\label{Equation_55}
\end{eqnarray}
and $g$ is the interaction constant, given by: $g=\pi\beta\alpha^{2}$.
The coefficients $\alpha_{i{\bf{k}}_{0}{\bf{q}}}$, $\beta_{i{\bf{k}}_{0}{\bf{q}}}$, $\gamma_{i{\bf{k}}_{0}{\bf{q}}}$ and $\delta_{i{\bf{k}}_{0}{\bf{q}}}$ are discussed in Appendix \ref{sec:Section_5}. The solution of the system in Eq.(\ref{Equation_54}) provides us with the numerical values and behavior of important physical quantities in the system, such as the chemical potential $\mu$, the average charge density imbalance $\delta{\bar{n}}$, the excitonic order parameter $\Delta_{\sigma}$, and the superconducting order parameter $\Delta_{{\rm s}\sigma}$. 
%
\subsection{\label{sec:Section_2_4} The orbitals in the ${\bf{k}}$-space}
%

Here, briefly, we will summarize our concept of possible spin-triplet superconductivity by just showing an important criteria for the establishment of the large orbitals in the $\bf{k}$-orbitals as the manifestation of the overall condensate state, as it is oftenly discussed in the context of superconducting state and order. In general the $p$-pairing between the electrons with the parallel spins costs much in the context of the energy in the $\bf{k}$-space. When the electrons start the exchanging process when they get paired with parallel spins the Fermi surface opens at that pairing places leading to the possibility to evolve the other electrons below the Fermi level in that pairing procedure. So, in the system and overall process starts that is inflamental to lead to the macroscopic formation of the condensate state in the Cu-based systems.     
It is well known that the binding energy of $f$-electrons in the Cu-based square lattice could be calculated via the standard Hydrogen-like formula for the energies \cite{cite_28, cite_29, cite_30} of the electronic orbitals at temperatures $T<160 K$ . For the case of copper we have $n=3$ and 
\begin{eqnarray}
E_{n=3}=-\frac{\rm{Ry}}{n^{2}}=-1.511 \ \  \text{eV},
\label{Equation_56}
\end{eqnarray}  
where $\rm {Ry}=13.6$ eV. We will treat the binding as the elastic process, having the influence of copper-medium on the last 4s$^{1}$ electrons which are plausible to for the pairing states. On the other hand that energy could be equalized to the formula of the sping energy in the elastic medium as
\begin{eqnarray}  
E_{n=3}=\varepsilon_{s}=\frac{\kappa_{\rm Cu}{R^{2}}}{2},
\label{Equation_57}
\end{eqnarray}
where $\kappa_{\rm Cu}$ is the sping constant of Cooper. The vector-variable $\bf{R}$ is the distance from the nucleus till the situation of the 4s$^{1}$ electron in the copper element electronic configuration. We have for the experimental value of the spring constant for copper the following value: $\kappa_{\rm Cu}=3.4931$ \cite{cite_31}. Then, from Eq.(\ref{Equation_57}), we can calculate the radii for the 4s$^{1}$ electron that is localized at the atomic site position of Cooper lattice in the insulating regime: 
\begin{eqnarray}     
R=\sqrt{\frac{2\varepsilon_{\rm f}}{\kappa_{\rm Cu}}}=2.79048 \ \ \ \rm{\AA}.
\label{Equation_58}
\end{eqnarray}
The corresponding length of the wave vector ${\bf{k}}$ of the 4s$^{1}$ electrons is of order of 
\begin{eqnarray}  
|{\bf{k}}|=\frac{2\pi}{R}=2.25\times 10^{8}  \ \ \ \rm{cm^{-1}}. 
\label{Equation_59}
\end{eqnarray}
On the other hand the Fermi level in our system is strongly dependent on the complicated interaction process in the system, even at the very low temperatures. We will consider the temperatures as high as $T=200$ K (see in panel (a), Fig.~\ref{fig:Fig_3}). At that temperature the reduced dimensionless parameter $T/\epsilon^{\left({\bf{q}}_{0}\right)}=0.465$. At the half-filling regime, with the inverse filling factor $\kappa=0.5$, the chemical potential in the system $\mu=-6.5\epsilon^{\left({\bf{q}}_{0}\right)}=-0.2405$ eV. Then, we can calculate the values of the wave vector corresponding to the chemical potential at $T=200$ K. We have
\begin{eqnarray}     
\mu=\frac{\kappa_{\rm Cu}\lambda^{2}}{2}=\frac{2\pi^{2}\kappa_{\rm Cu}}{{\bf{k}}^{2}_{\mu}}.
\label{Equation_60}
\end{eqnarray}
We obtain
\begin{eqnarray}     
|{\bf{k}}_{\mu}|=16.92 \times 10^{8} \rm{cm^{-1}}. 
\label{Equation_61}
\end{eqnarray}
The important next step is to calculate the wave vector corresponding to the elementary excitation $\epsilon^{\left({\bf{q}}_{0}\right)}=0.037$ eV, assumed at the begining of this paper (see in \ref{Equation_9}, on page 3). Indeed, we have
\begin{eqnarray} 
\epsilon^{\left({\bf{q}}_{0}\right)}=\frac{\kappa_{\rm Cu}\lambda^{2}_{{\bf{q}}_{0}}}{2}.
\label{Equation_62}
\end{eqnarray}
We obtain $\lambda_{{\bf{q}}_{0}}=0.145$ $\AA$. 
\begin{eqnarray}
|{\bf{k}}_{{\bf{q}}_{0}}|=\frac{2\pi}{\lambda_{{\bf{q}}_{0}}}=43.31 \times 10^{8} \rm{cm^{-1}} .
\label{Equation_63}
\end{eqnarray}
We see that the wave vector ${\bf{k}}_{{\bf{q}}_{0}}$ of the elementary excitation $\epsilon^{\left({\bf{q}}_{0}\right)}$, is much larger than the wave vector permitted for any quasiparticle excitation in the system. In other words, we have a threshold value ${\bf{k}}_{{\bf{q}}_{0}}$ and only the excitations with wave vectors ${\bf{k}}>{\bf{k}}_{{\bf{q}}_{0}}$ are possible in the system. Thus, the condition of establishment of the macroscopic superconducting condensate state in the system of square lattice copper system is well satisfied. Indeed, we have ${\bf{k}}_{{\bf{q}}_{0}}>{\bf{k}}_{\mu}$ and the electrons could pass au del\`a of Fermi surface, and, therefore, lead to the enlargement of the electronic orbitals in the ${\bf{k}}$-space. This is the one of the unique cases when the macroscopic changes in the system have the impact on the wave guidance in the reciprocal space and modify the topology of the electronic orbitals. This result is consistent with the suggestion in Ref.\onlinecite{cite_32, cite_33} about the enlargement of the electronic orbitals when the order parameter takes on a finite value.      
%
\section{\label{sec:Section_3} Numerical results and discussion}
\begin{figure}
	\includegraphics[scale=0.5]{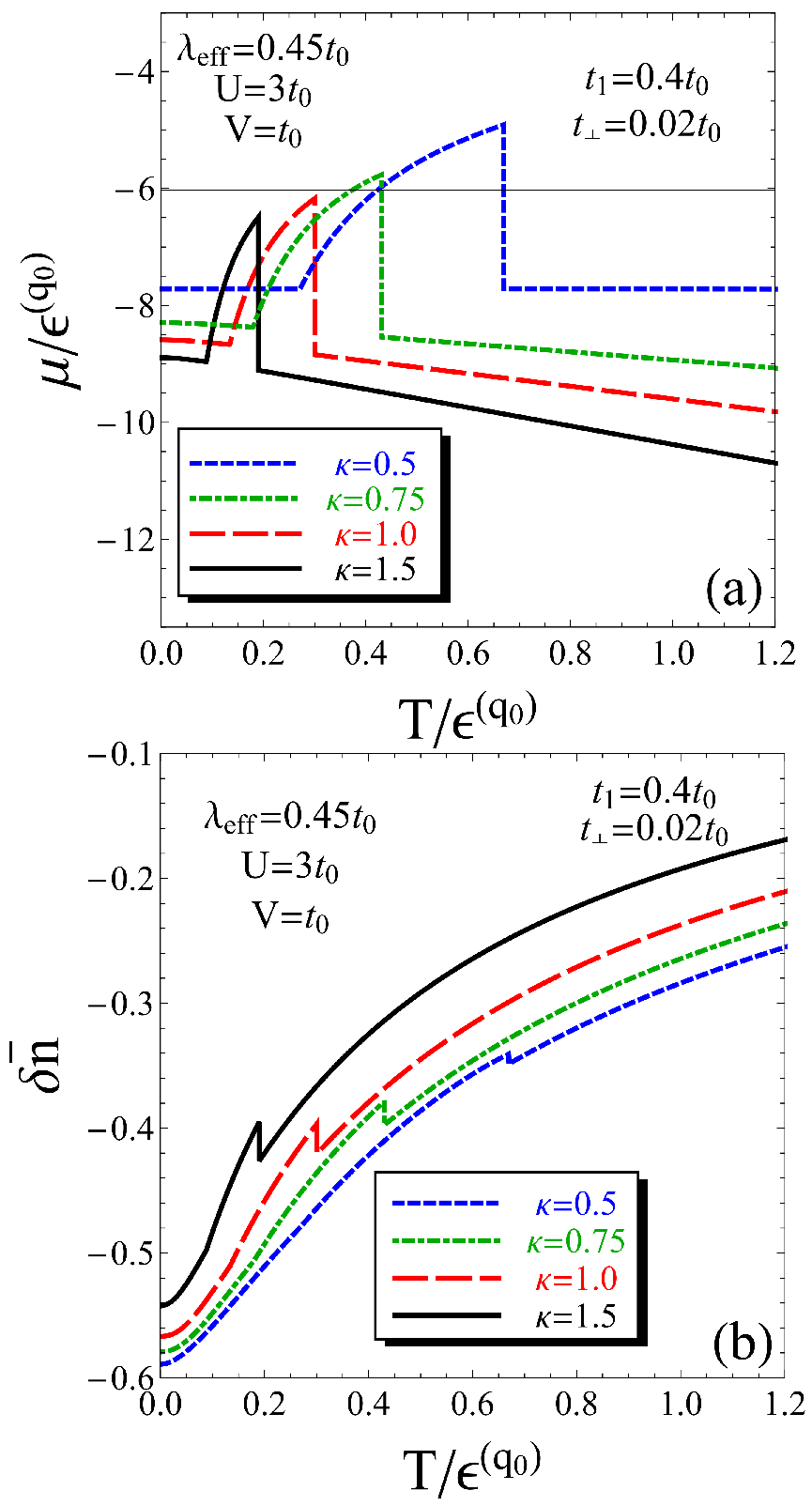}
	\caption{\label{fig:Fig_3}(Color online) The numerical solution of the system of self-consistent equations in Eq.(\ref{Equation_54}). The temperature dependence is shown for the chemical potential $\mu$ (see panel (a)) and the average interlayer charge density difference $\delta{\bar{n}}$ (see panel (b)). A large value of electron-phonon interaction parameter is considered with $\lambda_{\rm eff}=0.45t_{0}=0.135$ eV. The intralayer local Hubbard interaction parameter is set to $U=3t_{0}$, and the external gate potential is fixed at $V=t_{0}$. The hopping amplitudes are set to $t_{1}=0.4t_{0}$ and $t_{\perp}=0.02t_{0}$, respectively. The interlayer Hubbard interaction potential is $W=0.1t_{0}$. The calculations are performed for different values of the inverse filling coefficient $\kappa$.}
\end{figure}

%
In Fig.~\ref{fig:Fig_3}, we present the numerical results for the chemical potential $\mu$ (see panel (a)) and the interlayer charge density difference $\delta{\bar{n}}$ (see panel (b)). The system of equations in Eq.(\ref{Equation_54}) has been solved at the point $|{\bf{k}}_{0}|=\pi/3$ in the first Brillouin zone, considering the ${\bf{q}}$-excitations in the interval ${\bf{q}}\in\left[0.9{\bf{k}}_{0},1.1{\bf{k}}_{0}\right]$. 
%
%
%
\begin{figure}
	\includegraphics[scale=0.5]{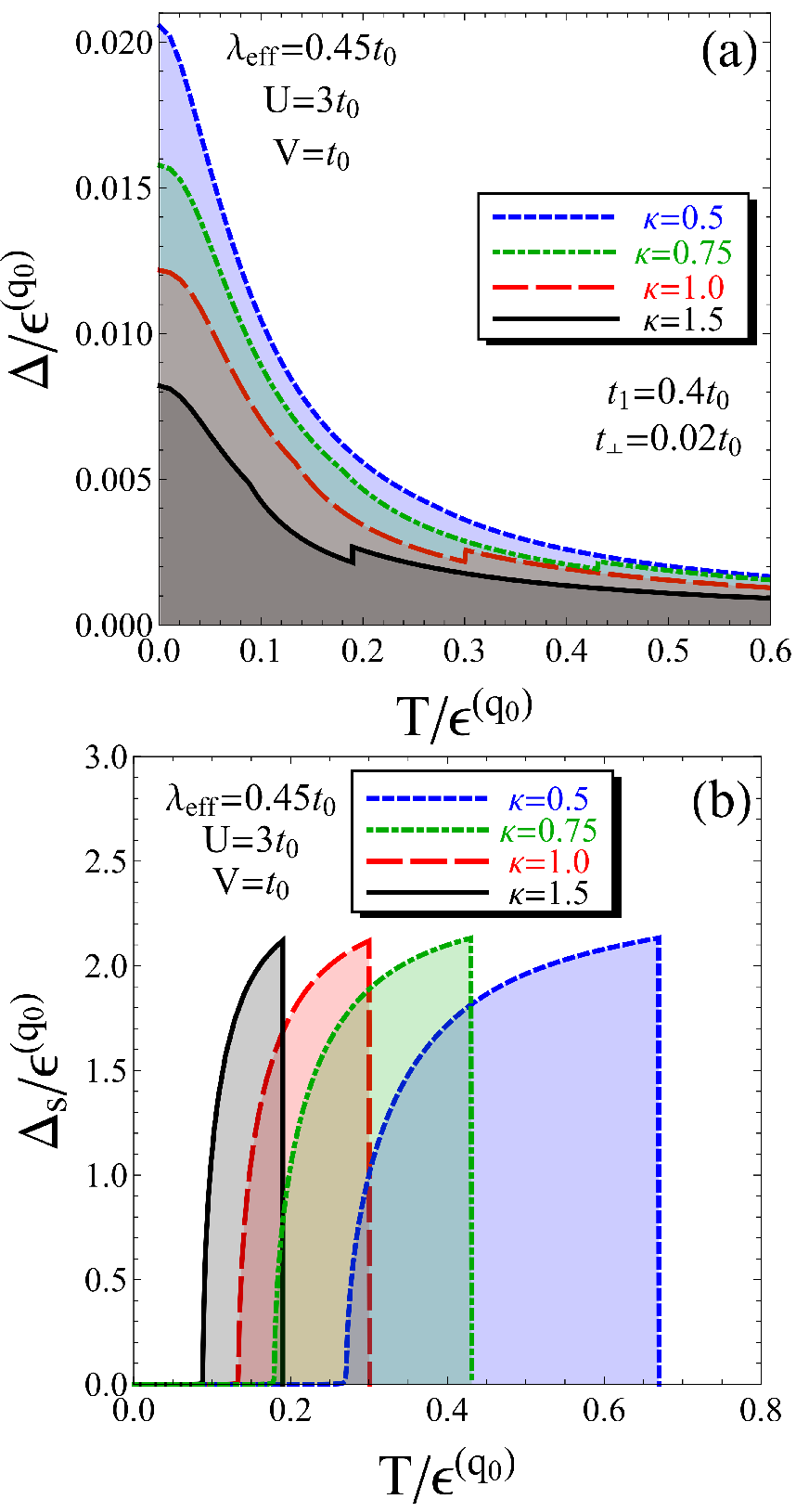}
	\caption{\label{fig:Fig_4}(Color online) The numerical solution of the system of self-consistent equations in Eq.(\ref{Equation_54}). The temperature dependence is shown for the excitonic order parameter $\Delta$ (see panel (a)) and the superconducting order parameter $\Delta_{s}$ (see panel (b)). A large values of electron-phonon interaction parameter is considered with $\lambda_{\rm eff}=0.45t_{0}$. The intralayer local Hubbard interaction parameter is set to $U=3t_{0}$, and the external gate potential is fixed at $V=t_{0}$. The hopping amplitudes are set to $t_{1}=0.4t_{0}$ and $t_{\perp}=0.02t_{0}$, respectively. The interlayer Hubbard interaction potential is $W=0.1t_{0}$. The calculations have been done for different values of the inverse filling coefficient $\kappa$.}
\end{figure} 
%
%
In Fig.~\ref{fig:Fig_4}, we present the self-consistent solution of the system in Eq.(\ref{Equation_54}) for the excitonic order parameter $\Delta$ (see panel (a)) and the superconducting order parameter $\Delta_{s}$ (see panel (b)).
The physical quantities in Figs.~\ref{fig:Fig_3} and ~\ref{fig:Fig_4} have been calculated as a function of temperature, for different values of the inverse filling coefficient $\kappa$, starting from the half-filling limit with $\kappa=0.5$ and extending to lower on-site occupancy limits (see different curves in Figs.~\ref{fig:Fig_3} and ~\ref{fig:Fig_4}). The effective electron-phonon interaction coefficient $\lambda_{\rm eff}$ has been fixed at $\lambda_{\rm eff}=0.45t_{0}=0.135$ eV, and the other interaction parameters have been set to the following values: intralayer Coulomb interaction $U=3t_{0}=0.9$ eV, external gate voltage: $V=t_{0}=0.3$ eV, and interlayer Coulomb interaction: $W=0.1t_{0}=0.03$ eV.
%
%
\begin{figure}
	\includegraphics[scale=0.5]{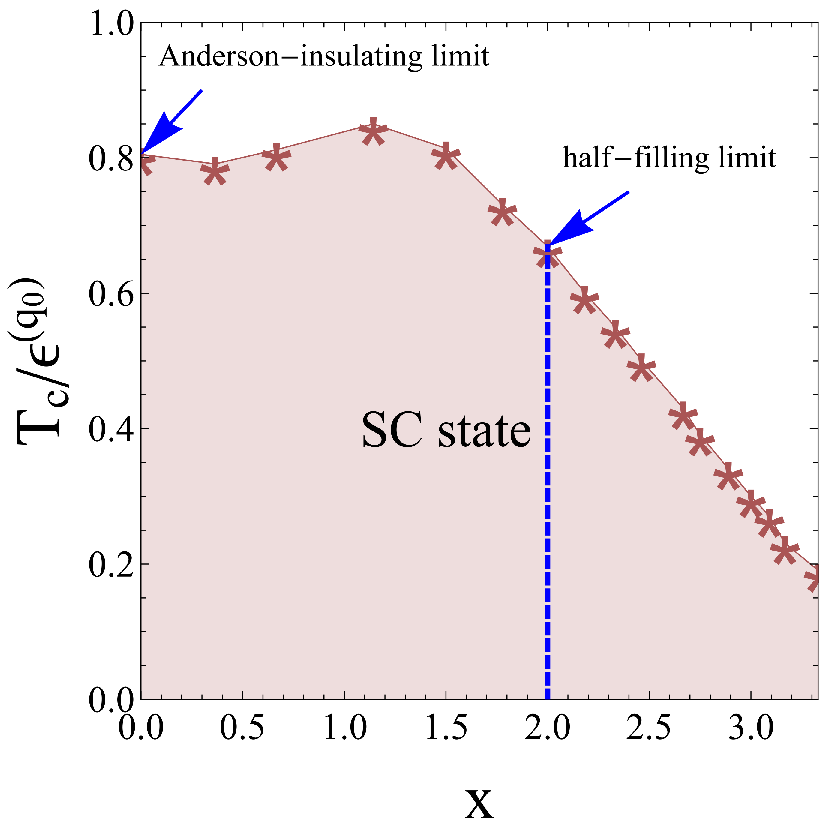}
	\caption{\label{fig:Fig_5}(Color online) The critical temperatures of the spin-triplet superconducting phase transition in a system of two metallic layers as a function of electron doping. The following values of the interaction parameters were chosen when evaluating the curve: electron-phonon interaction parameter: $\lambda_{\rm eff}=0.45t_{0}$, intralayer Hubbard interaction $U=3t_0$, interlayer Hubbard interaction parameter $W=0.1t_0$, and the external interlayer gate voltage $V=t_{0}$.The next-nearest neighbor (nnn) hopping was set at $t_{1}=0.4t_{0}$, and the interlayer hopping amplitude was fixed at $t_{\perp}=0.02t_{0}$.}
\end{figure} 
%
%
\begin{figure}
	\includegraphics[scale=0.5]{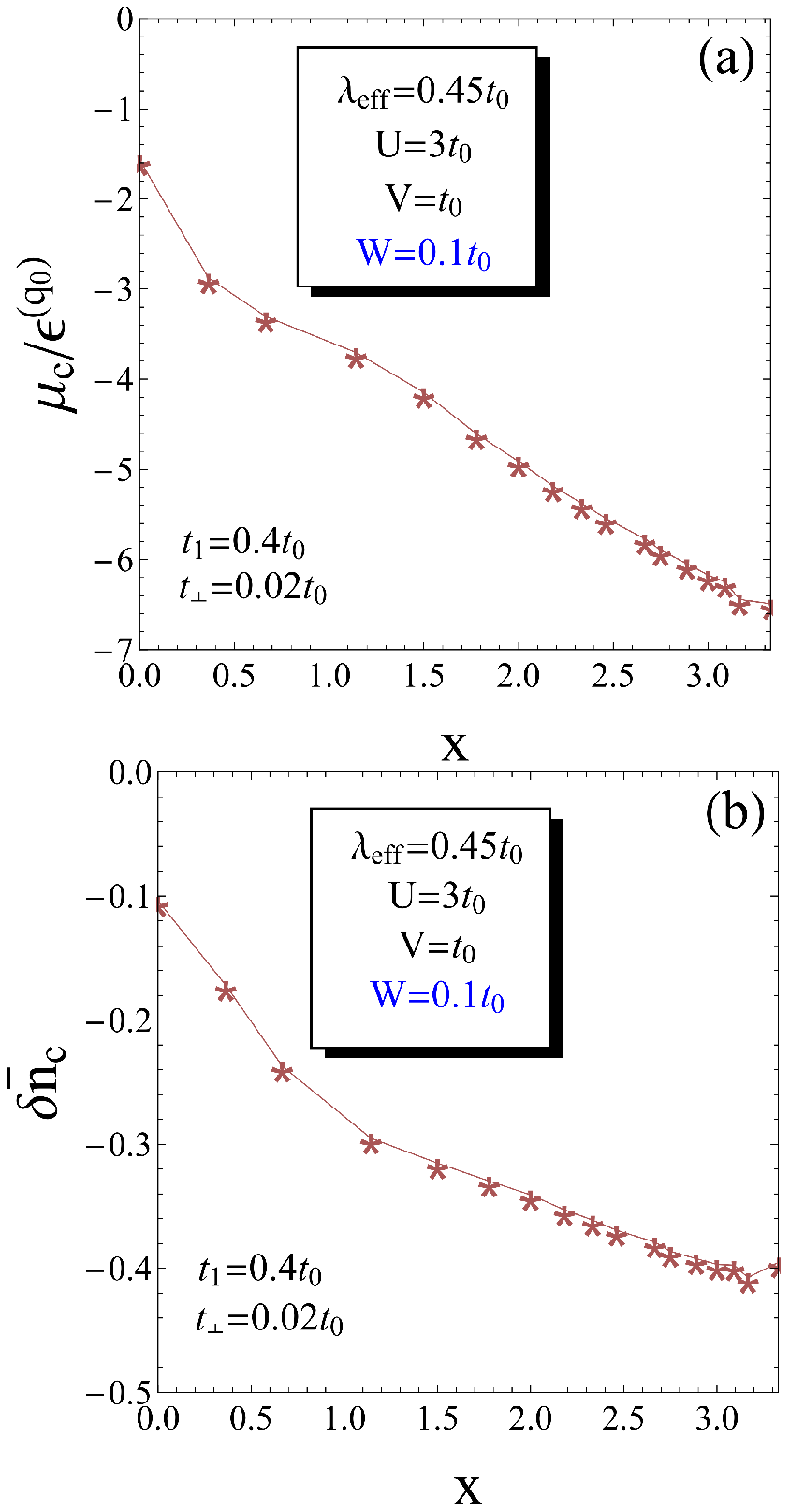}
	\caption{\label{fig:Fig_6}(Color online) The critical values of the chemical potential (see panel (a)) and the average electron density difference (see panel (b)) as functions of electron doping in the system. Each point on the curves corresponds to the respective value of the superconducting critical temperature $T_{c}$ (refer to Fig.~\ref{fig:Fig_5}). The effective electron-phonon interaction parameter is set at $\lambda_{\rm eff}=0.45t_{0}$, the next-nearest neighbor hopping is set at $t_{1}=0.4t_{0}$, and the interlayer hopping amplitude is fixed at value $t_{\perp}=0.02t_{0}$.}
\end{figure} 
%
%
\begin{figure}
	\includegraphics[scale=0.5]{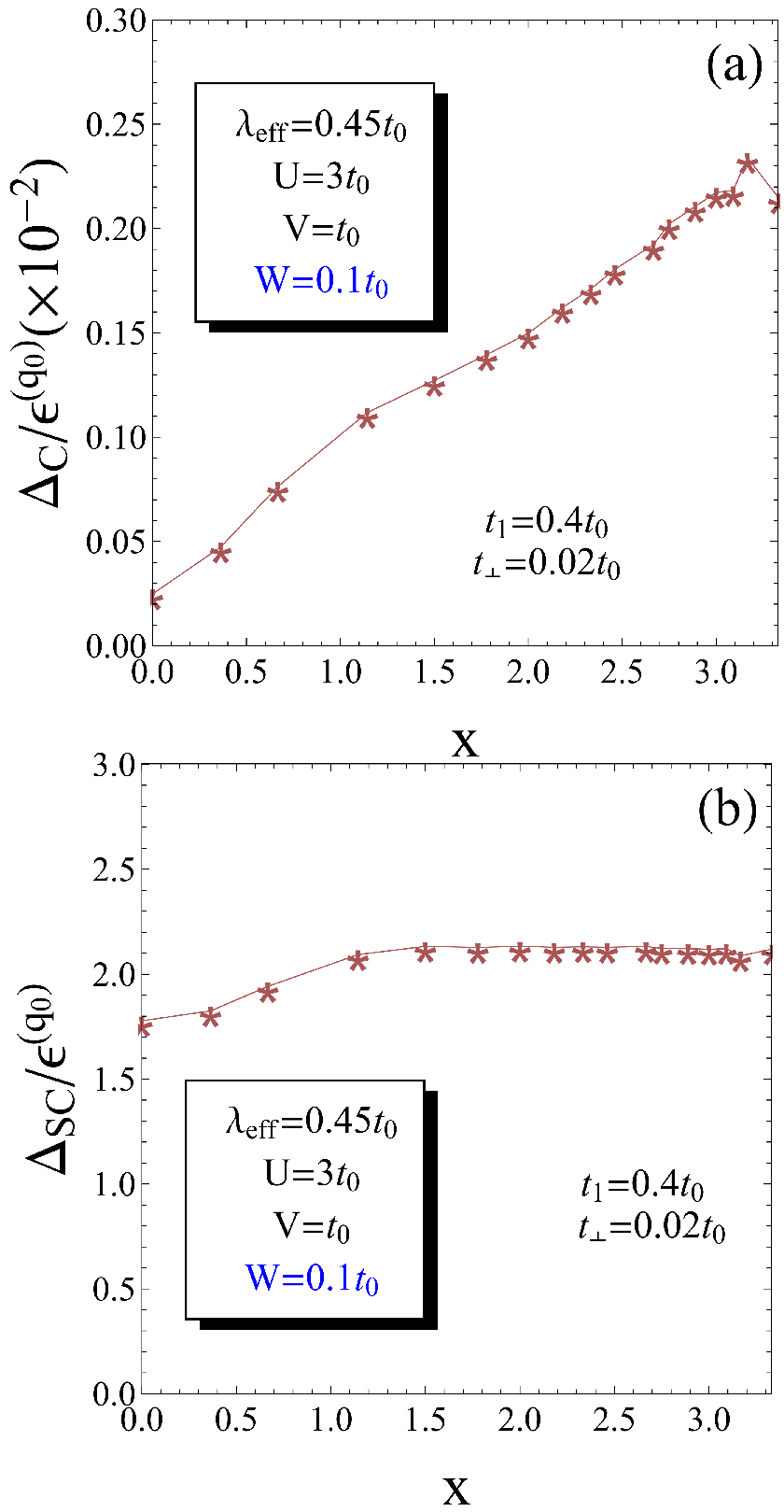}
	\caption{\label{fig:Fig_7}(Color online) The critical values of the excitonic (see panel (a)) and superconducting (see in panel (b)) order parameters as functions of electron doping in the system. Each point on the curves corresponds to the respective value of the superconducting critical temperature $T_{c}$ (refere to Fig.~\ref{fig:Fig_5}). The effective electron-phonon interaction parameter is set at $\lambda_{\rm eff}=0.45t_{0}$, the next-nearest neighbor hopping is at $t_{1}=0.4t_{0}$, and the interlayer hopping amplitude is fixed at $t_{\perp}=0.02t_{0}$.}
\end{figure} 
%
%
\begin{figure}
	\includegraphics[scale=0.5]{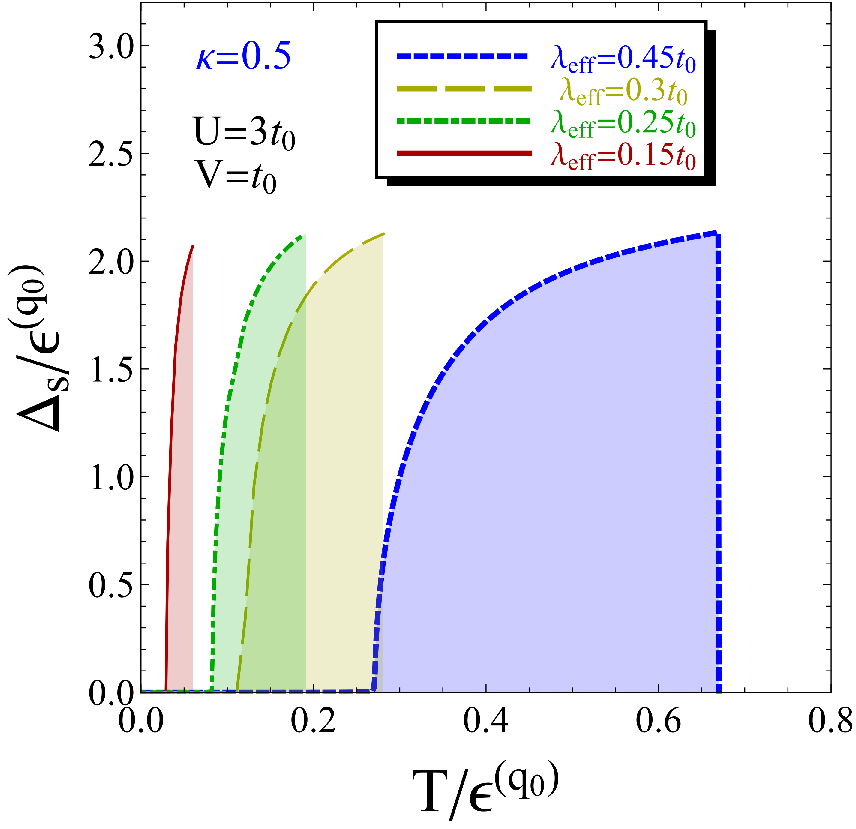}
	\caption{\label{fig:Fig_8}(Color online) The superconducting order parameters as a function of temperature. The calculations are performed for the case of half-filling with $\kappa=0.5$. Different values of the electron-phonon interaction parameter $\lambda_{\rm eff}$ are utilized. The following values of the interaction parameters were chosen when evaluating the curves: intralayer Hubbard interaction $U=3t_0$, interlayer Hubbard interaction parameter $W=0.1t_0$, and the external interlayer gate voltage $V=t_{0}$. The next-nearest neighbor (nnn) hopping is set at $t_{1}=0.4t_{0}$, and the interlayer hopping amplitude is fixed at value $t_{\perp}=0.02t_{0}$.}
\end{figure} 
%

The values used for the next-nearest neighbor hopping $t_{1}$ and interlayer hopping $t_{\perp}$ are shown in the figures. In particular, we have chosen: $t_{1}=0.4t_{0}>0$ (corresponding to electron doping in the system) \cite{cite_34} and $t_{\perp}=0.02t_{0}$. The system of self-consistent equations in Eq.(\ref{Equation_54}) has been solved by using the finite difference approximation method, which employs Newton's fast convergence algorithm \cite{cite_35}. In panel (a) of Fig.~\ref{fig:Fig_3}, we see that the chemical potential decreases in absolute value in the transition region, facilitating simultaneous single-particle excitations in the system. This leads, as we will see further, to the formation of a spin-triplet like superconducting state in the metallic bilayer system (as indicated by the black-square formations in Fig.~\ref{fig:Fig_2}). At higher temperatures, the chemical potential jumps to normal values, corresponding to the destruction of the superconducting order. The behavior of the average charge density difference $\delta{\bar{n}}$ is shown in panel (b) of Fig.~\ref{fig:Fig_3}. We observe that at the critical temperatures, when the chemical potential experiences drastic jumps, the function $\delta{\bar{n}}$ also changes significantly in its normal dependence on temperature $T$.     
The excitonic order parameter $\Delta$, calculated in panel (a), of Fig.~\ref{fig:Fig_4}, decreases with temperature until the critical temperature, at which it experiences slight jumps to higher values. The spin-triplet superconducting order parameter was calculated in panel (b), in Fig.~\ref{fig:Fig_4},  
and its behavior is consistent with the chemical potential curve shown in panel (a) of Fig.~\ref{fig:Fig_3}. We observe that the order parameter $\Delta_{\rm s}$ increases with temperature until the critical temperature $T_{\rm C}$, at which it vanishes, restoring the normal metal phase in the double-layer system. Additionally, we see that the transition temperature $T_{\rm C}$ is lowering as we move away from the half-filling limit $\kappa=0.5$ toward lower on-site occupancies with larger values of $\kappa$: $\kappa>0.5$. This behavior is consistent with the chemical potential curves in Fig.~\ref{fig:Fig_3}. Furthermore, in panel (a) of Fig.~\ref{fig:Fig_4}, we observe that the excitonic order parameter decreases as the parameter $\kappa$ increases.
The results in Fig.~\ref{fig:Fig_4} suggests that there is coexistence of the excitonic and superconducting states in the system over a large temperature range. The energy scales corresponding to superconducting order parameter are larger by two order of magnitude than the energy scales corresponding to the excitonic order parameter. 

Changing the values of the applied electric field potential $V$ just affect the amplitudes of the obtained curves, but the shapes of the curves still the same. Particularly, the large values of $V$ just decrease the amplitudes of the superconducting and excitonic order parameters in the system. Furthermore, when increasing the intralayer Coulomb interaction parameter $U$ leads to the slightly larger values of the excitonic and superconducting order parameters, meanwhile, affects considerably the  chemical potential and the average charge density imbalance between the layers leading the system towards the charge neutrality.  

After defining the doping in the system, as outlined in Eq.(\ref{Equation_42}), we can calculate the critical transition line corresponding to superconducting phase transition in the metallic bilayer system. This is presented in   
Fig.~\ref{fig:Fig_5}, where we have show the doping dependence of the transition critical temperature $T_{\rm C}$. As we observe, in the limit of low electron doping, the system is in the Anderson insulating state \cite{cite_36}, meaning that the atomic site positions are fully occupied by electrons. At a doping value of $x=2.0$, the system enters the half-filling regime. These limits are indicated by the blue arrows in Fig.~\ref{fig:Fig_5}. The values of the interaction parameters and the electron hopping amplitudes still the same as in Figs.~\ref{fig:Fig_3} and ~\ref{fig:Fig_4}. The calculated numerical values of the other physical quantities, corresponding to the transition curve in Fig.~\ref{fig:Fig_5} are shown in Figs.~\ref{fig:Fig_6} and ~\ref{fig:Fig_7} below.  
%
\begin{figure}
	\includegraphics[scale=0.3]{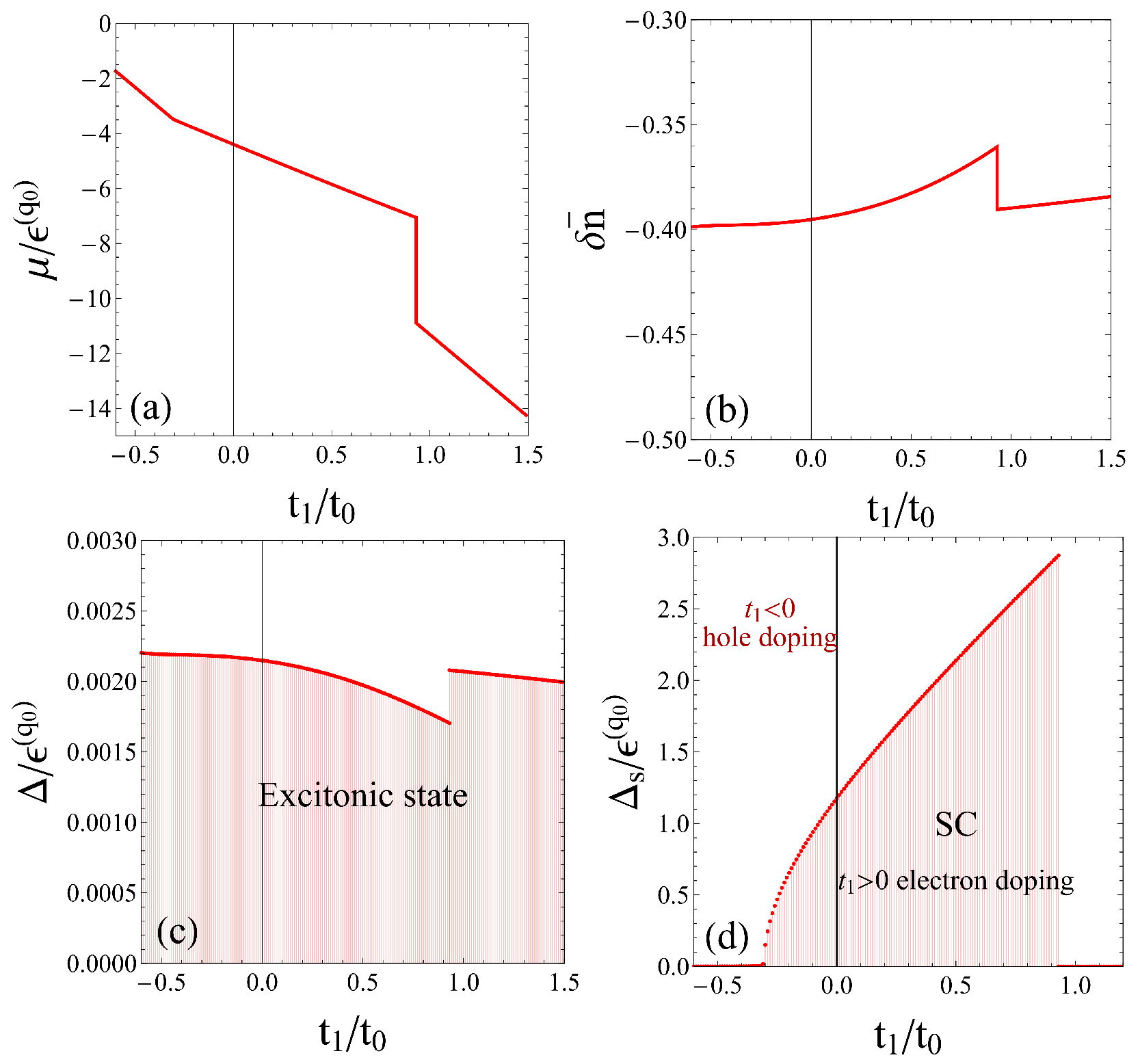}
	\caption{\label{fig:Fig_9}(Color online) The calculation of the physical quantities (see panels (a)-(d)) in the system as a function of the next next nearest neighbor (nnn) hopping amplitude $t_{1}$, for the case of half-filling with $\kappa=0.5$. The calculations are performed at $T=0.5\epsilon^{\left({\bf{q}}_{0}\right)}=214.68$ K. The following values of the interaction parameters were chosen when evaluating the curves: intralayer Hubbard interaction $U=3t_0$, interlayer Hubbard interaction $W=0.1t_0$, external interlayer gate voltage $V=t_{0}$ and electron-phonon interaction parameter $\lambda_{\rm eff}=0.45t_{0}$. The interlayer hopping amplitude is fixed at $t_{\perp}=0.02t_{0}$.}
\end{figure} 
%

Furthermore, in Fig.~\ref{fig:Fig_8}, we show the temperature dependence of the superconducting order parameter for different values of the electron-phonon interaction $\lambda_{\rm eff}$. The calculations in Fig.~\ref{fig:Fig_8} are performed for the case of half-filling with $\kappa=0.5$, which corresponds to a doping value of $x=2.0$. The other physical parameters are chosen as $U=3t_{0}$, $V=t_{0}$ and $W=0.1t_{0}$. As seen in Fig.~\ref{fig:Fig_8}, for large values of $\lambda_{\rm eff}$, the superconducting transition region is quite broad. When the electron-phonon interaction parameter is reduced, this transition region shifts to lower temperatures and also narrows in width. For example, the dark red curve in Fig.~\ref{fig:Fig_8}, corresponds to the low -$\lambda_{\rm eff}$ limit with $\lambda_{\rm eff}=0.15t_{0}$, where we find the superconducting transition critical temperature $T_{\rm C}=0.06\epsilon^{\left({\bf{q}}_{0}\right)}=25.76$ K. In contrast, the critical temperature corresponding to the case $\lambda_{\rm eff}=0.45t_{0}$ is $T_{\rm C}=0.6692\epsilon^{\left({\bf{q}}_{0}\right)}=287.3$ K.  
   
%
\begin{figure}
	\includegraphics[scale=0.3]{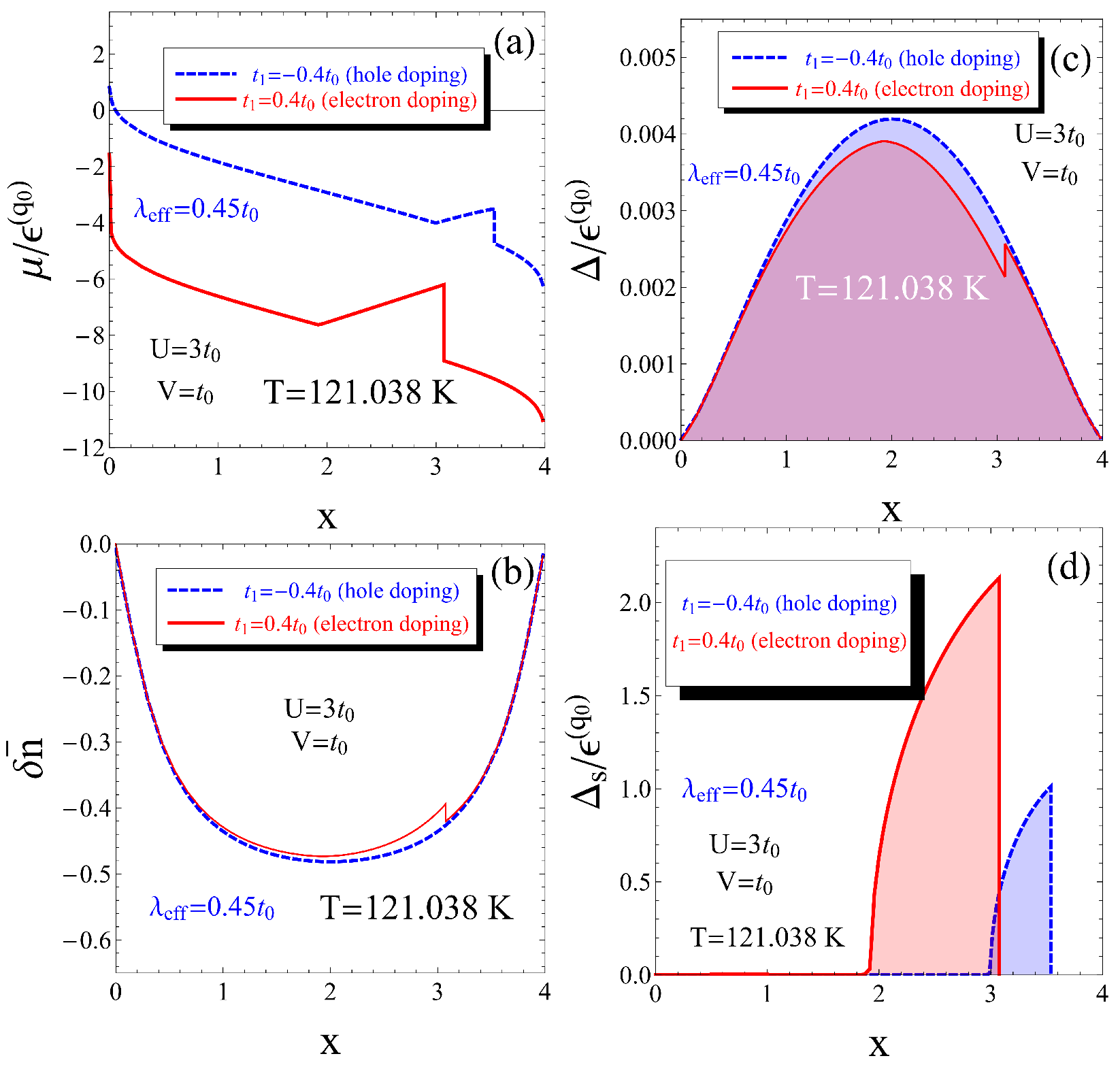}
	\caption{\label{fig:Fig_10}(Color online) The calculation of the physical quantities (see panels (a)-(d)) in the system as a function of doping. The calculations are performed at $T=121.038$ K. Two different values of the nnn hopping amplitudes are considered in Fig.~\ref{fig:Fig_10}, namely, $t_{1}=0.4t_{0}=0.12$ eV, corresponding to electron doping (see the red curve), and $t_{1}=-0.4t_{0}=-0.12$ eV for hole doping in the system (see the blue curve). The following values of the interaction parameters were chosen when evaluating the curves: intralayer Hubbard interaction $U=3t_0$, interlayer Hubbard interaction $W=0.1t_0$, external interlayer gate voltage $V=t_{0}$ and electron-phonon interaction parameter $\lambda_{\rm eff}=0.45t_{0}$. The interlayer hopping amplitude is fixed at $t_{\perp}=0.02t_{0}$.}
\end{figure} 
%
%
In Fig.~\ref{fig:Fig_9}, we solve the system of equations in Eq.(\ref{Equation_54}), and the results are presented as a function of the next-nearest neighbors hopping amplitude $t_{1}$. We observe that the chemical potential in panel (a) in Fig.~\ref{fig:Fig_9} experiences a drastic jump to larger (in absolute values) values at a critical value of the hopping amplitude $t_1=0.93t_{0}$. At this same value, other calculated physical quantities also exhibit significant changes: the average charge density difference $\delta{\bar{n}}$ is shown in panel (b), the excitonic order parameter is presented in (c), and the superconducting order parameter is depicted in panel (d). Both electron and hole states are illustrated in Fig.~\ref{fig:Fig_9}. In panels (c) and (d), in Fig.~\ref{fig:Fig_9}, we see that the excitonic and superconducting order parameters coexist for all values of $t_1$ in the interval $t_{1}\in\left[-0.6t_{0}, 1.5t_{0}\right]$. The superconducting order parameter vanishes in the negative region of the next-nearest neighbors hopping (which corresponds to hole doping in the system), while for positive values of $t_{1}$ it increases continuously with $t_1$ until reaching a maximum value of $\Delta_{\rm s}=106$ meV at $t_1=0.93t_0$.  
 
It is interesting to note a characteristic feature of the double-layer metallic bilayer in all calculation regimes considered in Figs.~\ref{fig:Fig_3}-~\ref{fig:Fig_9}: it works as an electron-hole bilayer. Specifically, this means that the lower layer $\ell=1$ is populated with more electrons than the top layer $\ell=2$, i.e., $\bar{n}_{2}<\bar{n}_{1}$.  

In Fig.~\ref{fig:Fig_10}, we shown the doping dependence of the same quantities at given temperature of $T=0.281t_{0}=121.038$ K. Two different values of the next next nearest neighbor hopping are considered in Fig.~\ref{fig:Fig_10}, namely, $t_{1}=0.4t_{0}=0.12$ eV, corresponding to electron doping, and $t_{1}=-0.4t_{0}=-0.12$ eV for hole doping in the system. The effective electron-phonon interaction parameter is fixed at $\lambda_{\rm eff}=0.45t_{0}=135$ meV, while the other tuning parameters are set as follows: $U=3t_{0}$, $V=t_{0}$, and $W=0.1t_{0}$. 

We observe in panel (a), in Fig.~\ref{fig:Fig_10}, that the absolute value of the chemical potential corresponding to electron doping (see the solid red curve) is lower than the absolute value of the chemical potential for hole doping (see the dashed blue curve). This indicates that single-particle excitations are favored more in the case of hole doping. For example, the excitonic gap parameter (see panel (b) in Fig.~\ref{fig:Fig_10}) is higher for $t_{1}=-0.4t_{0}$. The maximum value of the spin-singlet excitonic gap parameter occurs at half-filling (when $x=2.0$) for both cases. Specifically, we have $\Delta_{\rm max}=0.003891\epsilon^{\left({\bf{q}}_{0}\right)}=14.3$ meV for electron doping and $\Delta_{\rm max}=0.004196\epsilon^{\left({\bf{q}}_{0}\right)}=15.5$ meV for hole doping. 

An interesting behavior is observed in panel (c) of Fig.~\ref{fig:Fig_10} for the average charge density difference between the layers as a function of doping. For both considered cases, this function starts at $\delta{\bar{n}}=0$ at the Anderson localization limit indicating charge neutrality in the complete system and returns again to the charge neutrality value far away from the half-filling limit. For all other values of the electron (or hole) doping parameter, we have $\delta{\bar{n}}<0$, which is consistent with the results obtained above, indicating that the metallic bilayer behaves as an electron-hole bilayer (with electrons in the lower layer $\ell=1$ and holes in the upper layer $\ell=2$) type throughout an entire parameter range. 

In panel (d) of Fig.~\ref{fig:Fig_10}, we present the results for the superconducting gap parameter as a function of doping in the system. The values of the other tuning parameters remain the same as previously adopted. The critical temperature of the spin-triplet superconducting order parameter is higher in the case of hole doping (see the blue plot in panel (d)), while its magnitude is gradually greater (nearly twice) for the case of electron doping in the system (see the red plot in panel (d)). Moreover, the superconducting transition region is broader in the case of electron doping.            
%
\section{\label{sec:Section_4} Concluding remarks}
%
In this paper, we proposed an alternative mechanism for spin-triplet superconductivity in a double-layer square layer system, serving as a prototype model for the cuprate high-temperature superconductors (high-$T_C$). We address the electron-phonon interaction in terms of the total electron energy modification due to lattice distortion. On-site Hubbard interaction potentials are included in the calculations and the tight-binding component of the model- comprised of nearest neighbors, next-nearest neighbors, and inter-layer hopping terms- describes the electron hopping processes between different square lattice sites (within and between the layers). Before applying the formalism, we present the electron excitation factorization procedure, where a single-electron excitation with a given mode is represented as a product of two, simultaneous, single-particle excitations. We construct Nambu spinors for the problem at hand and solve the system of equations for a series of important physical quantities, including the chemical potential, average charge density imbalance between layers, spin- singlet excitonic pairing order parameter, and spin-triplet superconducting pairing gap function. Our calculations reveal the coexistence of excitonic and superconducting order parameters throughout the entire temperature transition region of the superconducting gap for the all parameter regimes. We calculate the temperature dependence of the aforementioned quantities for different electron doping regimes. The superconducting transition region is largest at half-filling, meanwhereas, this region narrows for fractional fillings (larger dopings). Consequently, the behavior of the chemical potential changes significantly in the superconducting transition region, while the average charge density difference between layers varies slowly as the system transitions to the normal metallic state. We construct a critical phase transition diagram for the model, in which the superconducting critical temperature is calculated as a function of doping, beginning from the Mott insulating limit and progressing through half-filling. Our results indicate that the superconducting transition critical temperature decreases dramatically at higher doping levels. 

We find the temperature dependence of the superconducting order parameter for a range of electron-phonon interaction parameter, discovering that the broadest spectrum of the superconducting phase transition occurs for larger values of this parameter. Additionally, we identify all possible values of the next-nearest neighbor hopping integral that can lead to a superconducting phase transition. 

The presented results in this manuscript are unique in the essence of the theoretical aspect, and hold considerable interest for the researchers working with strongly correlated electrons and contemporary aspects of high temperature superconductivity. The theory evaluated in the paper encloses, perhaps for the first time, the unresolved mechanism of superconductivity in high-$T_C$ superconductors and is adaptable to align with experimental regimes. A more comprehensive understanding remains invaluable, and further research is warranted to explore these intriguing phenomena. Just, more force costs nothing ...

\appendix

%
\section{\label{sec:Section_5} Self-consistent equations}
%
Here, we write the system of self-consistent equations at point ${\bf{k}}={\bf{q}}_{0}$ 
\begin{widetext}
\begin{eqnarray}
\frac{1}{2N_{\rm ph}}\sum_{{\bf{q}}}\sum_{\ell\nu_{n}\sigma}\left[\left\langle \bar{a}_{\ell\sigma}\left({\bf{k}}_{0}+\frac{{\bf{q}}}{2},\nu_{n}\right){a}_{\ell\sigma}\left({\bf{k}}_{0}+\frac{{\bf{q}}}{2},\nu_{n}\right)\right\rangle+\left\langle \bar{a}_{\ell\sigma}\left({\bf{k}}_{0}-\frac{{\bf{q}}}{2},\nu_{n}\right){a}_{\ell\sigma}\left({\bf{k}}_{0}-\frac{{\bf{q}}}{2},\nu_{n}\right)\right\rangle\right]=\frac{1}{\kappa},
\nonumber\\
\frac{1}{2N_{\rm ph}}\sum_{{\bf{q}}}\sum_{\ell\nu_{n}\sigma}\left(-1\right)^{\ell}\left[\left\langle \bar{a}_{\ell\sigma}\left({\bf{k}}_{0}+\frac{{\bf{q}}}{2},\nu_{n}\right){a}_{\ell\sigma}\left({\bf{k}}_{0}+\frac{{\bf{q}}}{2},\nu_{n}\right)\right\rangle+\left\langle \bar{a}_{\ell\sigma}\left({\bf{k}}_{0}-\frac{{\bf{q}}}{2},\nu_{n}\right){a}_{\ell\sigma}\left({\bf{k}}_{0}-\frac{{\bf{q}}}{2},\nu_{n}\right)\right\rangle\right]=\delta{\bar{n}},
\nonumber\\
\Delta=\frac{W}{2N_{\rm ph}}\sum_{{\bf{q}}}\sum_{\nu_{n}\sigma}\left[\left\langle \bar{a}_{2\sigma}\left({\bf{k}}_{0}+\frac{{\bf{q}}}{2},\nu_{n}\right){a}_{1\sigma}\left({\bf{k}}_{0}+\frac{{\bf{q}}}{2},\nu_{n}\right)\right\rangle+\left\langle \bar{a}_{2\sigma}\left({\bf{k}}_{0}-\frac{{\bf{q}}}{2},\nu_{n}\right){a}_{1\sigma}\left({\bf{k}}_{0}-\frac{{\bf{q}}}{2},\nu_{n}\right)\right\rangle+\right.
\nonumber\\
\left.+\left\langle \bar{a}_{1\sigma}\left({\bf{k}}_{0}+\frac{{\bf{q}}}{2},\nu_{n}\right){a}_{2\sigma}\left({\bf{k}}_{0}+\frac{{\bf{q}}}{2},\nu_{n}\right)\right\rangle+\left\langle \bar{a}_{1\sigma}\left({\bf{k}}_{0}-\frac{{\bf{q}}}{2},\nu_{n}\right){a}_{2\sigma}\left({\bf{k}}_{0}-\frac{{\bf{q}}}{2},\nu_{n}\right)\right\rangle\right],
\nonumber\\
\Delta_{\rm s}=\frac{g}{4N_{\rm ph}}\sum_{{\bf{q}}}\sum_{\nu_{n}\sigma}\left[\left\langle \bar{a}_{1\sigma}\left({\bf{k}}_{0}-\frac{{\bf{q}}}{2},\nu_{n}\right){a}_{1\sigma}\left({\bf{k}}_{0}+\frac{{\bf{q}}}{2},\nu_{n}\right)\right\rangle+\left\langle \bar{a}_{1\sigma}\left({\bf{k}}_{0}+\frac{{\bf{q}}}{2},\nu_{n}\right){a}_{1\sigma}\left({\bf{k}}_{0}-\frac{{\bf{q}}}{2},\nu_{n}\right)\right\rangle+\right.
\nonumber\\
\left.+\left\langle \bar{a}_{2\sigma}\left({\bf{k}}_{0}-\frac{{\bf{q}}}{2},\nu_{n}\right){a}_{2\sigma}\left({\bf{k}}_{0}+\frac{{\bf{q}}}{2},\nu_{n}\right)\right\rangle+\left\langle \bar{a}_{2\sigma}\left({\bf{k}}_{0}+\frac{{\bf{q}}}{2},\nu_{n}\right){a}_{2\sigma}\left({\bf{k}}_{0}-\frac{{\bf{q}}}{2},\nu_{n}\right)\right\rangle\right].
\label{Equation_A_1}
\end{eqnarray}
\end{widetext}
The statistical averages entering under the sums in each equation in the system of equations in Eq.(\ref{Equation_A_1}) could be evaluated after introducing the external source terms in the expression of the partition function of the system (and after the appropriate Hubbard-Stratonovich integration over Nambu spinors introduced in Eqs.(\ref{Equation_45}) and (\ref{Equation_46}). We introduce auxiliary decoupling fields in the form analogue to Nambu vectors ${\Psi}_{{\bf{k}}{\bf{q}}\sigma}\left(\nu_{n}\right)$ and $\bar{\Psi}_{{\bf{k}}{\bf{q}}\sigma}\left(\nu_{n}\right)$ in Eqs.(\ref{Equation_45}) and (\ref{Equation_46}), in the Section \ref{sec:Section_2_3}:

\begin{eqnarray} 
J_{{\bf{k}}_{0}{\bf{q}}\sigma}\left(\nu_{n}\right)=\left(
\begin{array}{crrrr}
\bar{J}_{a_{1}\sigma}\left({\bf{k}}_{0}-\frac{{\bf{q}}}{2},\nu_{n}\right)\\\\
{J}_{a_{1}\sigma}\left({\bf{k}}_{0}+\frac{{\bf{q}}}{2},\nu_{n}\right) \\\\
\bar{J}_{a_{2}\sigma}\left({\bf{k}}_{0}-\frac{{\bf{q}}}{2},\nu_{n}\right) \\\\
J_{a_{2}\sigma}\left({\bf{k}}_{0}+\frac{{\bf{q}}}{2},\nu_{n}\right) \\\\
\end{array}\right).
\label{Equation_A_2}
\end{eqnarray}
The corresponding complex conjugate spinor is denoted as 
\begin{widetext}
\begin{eqnarray} 
\bar{J}_{{\bf{k}}_{0}{\bf{q}}\sigma}\left(\nu_{n}\right)=\left({J}_{a_{1}\sigma}\left({\bf{k}}_{0}-\frac{{\bf{q}}}{2},\nu_{n}\right),\bar{J}_{a_{1}\sigma}\left({\bf{k}}_{0}+\frac{{\bf{q}}}{2},\nu_{n}\right), {J}_{a_{2}\sigma}\left({\bf{k}}_{0}-\frac{{\bf{q}}}{2},\nu_{n}\right),\bar{J}_{a_{2}\sigma}\left({\bf{k}}_{0}+\frac{{\bf{q}}}{2},\nu_{n}\right)\right).
\label{Equation_A_3}
\end{eqnarray}
\end{widetext}
The partition function of the system could be written as:
\begin{widetext}
\begin{eqnarray}
{\cal{Z}}&=&\int\left[{\cal{D}}\bar{\Psi}{\cal{D}}\Psi\right]e^{-\frac{1}{2}\sum_{{\bf{q}}\nu_{n},\sigma}\bar{\Psi}_{{\bf{k}}_{0}{\bf{q}}\sigma}\left(\nu_{n}\right)\beta{\cal{G}}^{-1}_{{\bf{k}}_{0}{\bf{q}}\sigma}\left(\nu_{n}\right){\Psi}_{{\bf{k}}_{0}{\bf{q}}\sigma}\left(\nu_{n}\right)}e^{\frac{1}{2}\sum_{{\bf{q}}\nu_{n},\sigma}\bar{J}_{{\bf{k}}_{0}{\bf{q}}\sigma}\left(\nu_{n}\right){\Psi}_{{\bf{k}}_{0}{\bf{q}}\sigma}\left(\nu_{n}\right)+\bar{\Psi}_{{\bf{k}}_{0}{\bf{q}}\sigma}\left(\nu_{n}\right){J}_{{\bf{k}}_{0}{\bf{q}}\sigma}\left(\nu_{n}\right)}
\nonumber\\
&&\approx e^{\frac{1}{2}\sum_{{\bf{q}}\nu_{n},\sigma}\bar{J}_{{\bf{k}}_{0}{\bf{q}}\sigma}\left(\nu_{n}\right){\cal{D}}_{{\bf{k}}_{0}{\bf{q}}\sigma}\left(\nu_{n}\right){J}_{{\bf{k}}_{0}{\bf{q}}\sigma}\left(\nu_{n}\right)},
\label{Equation_A_4}
\end{eqnarray}
\end{widetext}
where ${\cal{D}}_{{\bf{k}}_{0}{\bf{q}}\sigma}$ in the exponential in the right hand side in Eq.(\ref{Equation_A_4}) is the inverse of the matrix $\beta{\cal{G}}^{-1}_{{\bf{k}}_{0}{\bf{q}}\sigma}\left(\nu_{n}\right)$.
Then, we can calculate the averages under the sums in Eq.(\ref{Equation_A_1}). As an example, the first term, in the first equation, in Eq.(\ref{Equation_A_1}) could be calculated with the help of the functional differentiation techniques:
\begin{eqnarray}
&&\frac{\delta^{2}{\cal{Z}}}{\delta{\bar{J}_{a_1\sigma}}\left({\bf{k}}_{0}+\frac{{\bf{q}}}{2},\nu_n\right)\delta{J_{a_1\sigma}}\left({\bf{k}}_{0}+\frac{{\bf{q}}}{2},\nu_n\right)}=-\frac{1}{2}{\cal{D}}_{22{\bf{k}}_{0}{\bf{q}}\sigma}\left(\nu_n\right)
\nonumber\\
&&=-\frac{1}{4}\left\langle \bar{a}_{1\sigma}\left({\bf{k}}_{0}+\frac{{\bf{q}}}{2},\nu_{n}\right){a}_{1\sigma}\left({\bf{k}}_{0}+\frac{{\bf{q}}}{2},\nu_{n}\right)\right\rangle,
\label{Equation_A_5}
\end{eqnarray}
therefore, we get
\begin{eqnarray}
\left\langle \bar{a}_{1\sigma}\left({\bf{k}}_{0}+\frac{{\bf{q}}}{2},\nu_{n}\right){a}_{1\sigma}\left({\bf{k}}_{0}+\frac{{\bf{q}}}{2},\nu_{n}\right)\right\rangle=2{\cal{D}}_{22{\bf{k}}_{0}{\bf{q}}\sigma}\left(\nu_n\right).
\nonumber\\
\label{Equation_A_6}
\end{eqnarray}
Similarly, for the other averages in Eq.(\ref{Equation_A_1}) we get:
\begin{eqnarray}
\left\langle \bar{a}_{1\sigma}\left({\bf{k}}_{0}-\frac{{\bf{q}}}{2},\nu_{n}\right){a}_{1\sigma}\left({\bf{k}}_{0}-\frac{{\bf{q}}}{2},\nu_{n}\right)\right\rangle=-2{\cal{D}}_{11{\bf{k}}_{0}{\bf{q}}\sigma}\left(\nu_n\right),
\nonumber\\
\left\langle \bar{a}_{2\sigma}\left({\bf{k}}_{0}+\frac{{\bf{q}}}{2},\nu_{n}\right){a}_{2\sigma}\left({\bf{k}}_{0}+\frac{{\bf{q}}}{2},\nu_{n}\right)\right\rangle=2{\cal{D}}_{44{\bf{k}}_{0}{\bf{q}}\sigma}\left(\nu_n\right),
\nonumber\\
\left\langle \bar{a}_{2\sigma}\left({\bf{k}}_{0}-\frac{{\bf{q}}}{2},\nu_{n}\right){a}_{2\sigma}\left({\bf{k}}_{0}-\frac{{\bf{q}}}{2},\nu_{n}\right)\right\rangle=-2{\cal{D}}_{33{\bf{k}}_{0}{\bf{q}}\sigma}\left(\nu_n\right),
\nonumber\\
\left\langle \bar{a}_{2\sigma}\left({\bf{k}}_{0}+\frac{{\bf{q}}}{2},\nu_{n}\right){a}_{1\sigma}\left({\bf{k}}_{0}+\frac{{\bf{q}}}{2},\nu_{n}\right)\right\rangle=2{\cal{D}}_{24{\bf{k}}_{0}{\bf{q}}\sigma}\left(\nu_n\right),
\nonumber\\
\left\langle \bar{a}_{2\sigma}\left({\bf{k}}_{0}-\frac{{\bf{q}}}{2},\nu_{n}\right){a}_{1\sigma}\left({\bf{k}}_{0}-\frac{{\bf{q}}}{2},\nu_{n}\right)\right\rangle=-2{\cal{D}}_{31{\bf{k}}_{0}{\bf{q}}\sigma}\left(\nu_n\right),
\nonumber\\
\left\langle \bar{a}_{1\sigma}\left({\bf{k}}_{0}+\frac{{\bf{q}}}{2},\nu_{n}\right){a}_{2\sigma}\left({\bf{k}}_{0}+\frac{{\bf{q}}}{2},\nu_{n}\right)\right\rangle=2{\cal{D}}_{42{\bf{k}}_{0}{\bf{q}}\sigma}\left(\nu_n\right),
\nonumber\\
\left\langle \bar{a}_{1\sigma}\left({\bf{k}}_{0}-\frac{{\bf{q}}}{2},\nu_{n}\right){a}_{2\sigma}\left({\bf{k}}_{0}-\frac{{\bf{q}}}{2},\nu_{n}\right)\right\rangle=-2{\cal{D}}_{13{\bf{k}}_{0}{\bf{q}}\sigma}\left(\nu_n\right),
\nonumber\\
\left\langle {a}_{1\sigma}\left({\bf{k}}_{0}-\frac{{\bf{q}}}{2},\nu_{n}\right){a}_{1\sigma}\left({\bf{k}}_{0}+\frac{{\bf{q}}}{2},\nu_{n}\right)\right\rangle=2{\cal{D}}_{21{\bf{k}}_{0}{\bf{q}}\sigma}\left(\nu_n\right),
\nonumber\\
\left\langle \bar{a}_{1\sigma}\left({\bf{k}}_{0}+\frac{{\bf{q}}}{2},\nu_{n}\right)\bar{a}_{1\sigma}\left({\bf{k}}_{0}-\frac{{\bf{q}}}{2},\nu_{n}\right)\right\rangle=2{\cal{D}}_{12{\bf{k}}_{0}{\bf{q}}\sigma}\left(\nu_n\right),
\nonumber\\
\left\langle {a}_{2\sigma}\left({\bf{k}}_{0}-\frac{{\bf{q}}}{2},\nu_{n}\right){a}_{2\sigma}\left({\bf{k}}_{0}+\frac{{\bf{q}}}{2},\nu_{n}\right)\right\rangle=2{\cal{D}}_{43{\bf{k}}_{0}{\bf{q}}\sigma}\left(\nu_n\right),
\nonumber\\
\left\langle \bar{a}_{2\sigma}\left({\bf{k}}_{0}+\frac{{\bf{q}}}{2},\nu_{n}\right)\bar{a}_{2\sigma}\left({\bf{k}}_{0}-\frac{{\bf{q}}}{2},\nu_{n}\right)\right\rangle=2{\cal{D}}_{34{\bf{k}}_{0}{\bf{q}}\sigma}\left(\nu_n\right).
\nonumber\\
\label{Equation_A_7}
\end{eqnarray}
Furthermore, the system of equations, in terms of the inverse Green's function matrix will be rewritten as:
\begin{widetext}
\begin{eqnarray}
\frac{2}{N_{\rm ph}}\sum_{{\bf{q}}\nu_n,\sigma}\left({\cal{D}}_{22{\bf{k}}_{0}{\bf{q}}\sigma}\left(\nu_n\right)+{\cal{D}}_{44{\bf{k}}_{0}{\bf{q}}\sigma}\left(\nu_n\right)-{\cal{D}}_{11{\bf{k}}_{0}{\bf{q}}\sigma}\left(\nu_n\right)-{\cal{D}}_{32{\bf{k}}_{0}{\bf{q}}\sigma}\left(\nu_n\right)\right)=\frac{1}{\kappa},
\nonumber\\
\frac{2}{N_{\rm ph}}\sum_{{\bf{q}}\nu_n,\sigma}\left({\cal{D}}_{44{\bf{k}}_{0}{\bf{q}}\sigma}\left(\nu_n\right)-{\cal{D}}_{33{\bf{k}}_{0}{\bf{q}}\sigma}\left(\nu_n\right)-{\cal{D}}_{22{\bf{k}}_{0}{\bf{q}}\sigma}\left(\nu_n\right)+{\cal{D}}_{11{\bf{k}}_{0}{\bf{q}}\sigma}\left(\nu_n\right)\right)=\delta{\bar{n}},
\nonumber\\
\frac{W}{2N_{\rm ph}}\sum_{{\bf{q}}\nu_n,\sigma}\left({\cal{D}}_{24{\bf{k}}_{0}{\bf{q}}\sigma}\left(\nu_n\right)+{\cal{D}}_{42{\bf{k}}_{0}{\bf{q}}\sigma}\left(\nu_n\right)-{\cal{D}}_{31{\bf{k}}_{0}{\bf{q}}\sigma}\left(\nu_n\right)-{\cal{D}}_{13{\bf{k}}_{0}{\bf{q}}\sigma}\left(\nu_n\right)\right)=\Delta_{\sigma},
\nonumber\\
\frac{g}{2N_{\rm ph}}\sum_{{\bf{q}}\nu_n,\sigma}\left({\cal{D}}_{21{\bf{k}}_{0}{\bf{q}}\sigma}\left(\nu_n\right)+{\cal{D}}_{12{\bf{k}}_{0}{\bf{q}}\sigma}\left(\nu_n\right)+{\cal{D}}_{43{\bf{k}}_{0}{\bf{q}}\sigma}\left(\nu_n\right)+{\cal{D}}_{34{\bf{k}}_{0}{\bf{q}}\sigma}\left(\nu_n\right)\right)=\Delta_{\rm s}.
\label{Equation_A_8}
\end{eqnarray}
For the expressions, under the sums, in Eq.(\ref{Equation_A_8}), we have
\begin{eqnarray}
{\cal{D}}_{22{\bf{k}}_{0}{\bf{q}}\sigma}\left(\nu_n\right)+{\cal{D}}_{44{\bf{k}}_{0}{\bf{q}}\sigma}\left(\nu_n\right)-{\cal{D}}_{11{\bf{k}}_{0}{\bf{q}}\sigma}\left(\nu_n\right)-{\cal{D}}_{32{\bf{k}}_{0}{\bf{q}}\sigma}\left(\nu_n\right)=\frac{{\cal{P}}^{\left(3\right)}_{\mu}\left(i\nu_n\right)}{\det{\beta{{\cal{G}}^{-1}_{{\bf{k}}_{0}{\bf{q}}\sigma}\left(\nu_{n}\right)}}},
\nonumber\\
=\frac{1}{\beta}\frac{4\left(-i\nu_n\right)^{3}+\xi_{1}\left(-i\nu_n\right)^{2}+\xi_{2}\left(-i\nu_n\right)+\xi_{3}}{\prod^{4}_{i=1}\left(-i\nu_{n}-\epsilon_{i{\bf{k}}_{0}{\bf{q}}}\right)},
\nonumber\\
{\cal{D}}_{44{\bf{k}}_{0}{\bf{q}}\sigma}\left(\nu_n\right)-{\cal{D}}_{33{\bf{k}}_{0}{\bf{q}}\sigma}\left(\nu_n\right)-{\cal{D}}_{22{\bf{k}}_{0}{\bf{q}}\sigma}\left(\nu_n\right)+{\cal{D}}_{11{\bf{k}}_{0}{\bf{q}}\sigma}\left(\nu_n\right)=\frac{{\cal{P}}^{\left(3\right)}_{\delta{\bar{n}}}\left(i\nu_n\right)}{\det{\beta{{\cal{G}}^{-1}_{{\bf{k}}_{0}{\bf{q}}\sigma}\left(\nu_{n}\right)}}}
\nonumber\\
=\frac{1}{\beta}\frac{-4a\left(U,V\right)\left(-i\nu_n\right)^{2}+\xi_{4}\left(-i\nu_n\right)+\xi_{5}}{\prod^{4}_{i=1}\left(-i\nu_{n}-\epsilon_{i{\bf{k}}_{0}{\bf{q}}}\right)},
\nonumber\\
{\cal{D}}_{24{\bf{k}}_{0}{\bf{q}}\sigma}\left(\nu_n\right)+{\cal{D}}_{42{\bf{k}}_{0}{\bf{q}}\sigma}\left(\nu_n\right)-{\cal{D}}_{31{\bf{k}}_{0}{\bf{q}}\sigma}\left(\nu_n\right)-{\cal{D}}_{13{\bf{k}}_{0}{\bf{q}}\sigma}\left(\nu_n\right)=\frac{{\cal{P}}^{\left(2\right)}_{\Delta}\left(i\nu_n\right)}{\det{\beta{{\cal{G}}^{-1}_{{\bf{k}}_{0}{\bf{q}}\sigma}\left(\nu_{n}\right)}}}
\nonumber\\
=\frac{1}{\beta}\frac{\xi_{6}\left(-i\nu_n\right)^{2}+\xi_{7}\left(-i\nu_n\right)+\xi_{8}}{\prod^{4}_{i=1}\left(-i\nu_{n}-\epsilon_{i{\bf{k}}_{0}{\bf{q}}}\right)},
\nonumber\\
{\cal{D}}_{21{\bf{k}}_{0}{\bf{q}}\sigma}\left(\nu_n\right)+{\cal{D}}_{12{\bf{k}}_{0}{\bf{q}}\sigma}\left(\nu_n\right)+{\cal{D}}_{43{\bf{k}}_{0}{\bf{q}}\sigma}\left(\nu_n\right)+{\cal{D}}_{34{\bf{k}}_{0}{\bf{q}}\sigma}\left(\nu_n\right)=\frac{{\cal{P}}^{\left(2\right)}_{\Delta_{\rm s}}\left(i\nu_n\right)}{\det{\beta{{\cal{G}}^{-1}_{{\bf{k}}_{0}{\bf{q}}\sigma}\left(\nu_{n}\right)}}}
\nonumber\\
=\frac{1}{\beta}\frac{\xi_{9}\left(-i\nu_n\right)^{2}+\xi_{10}\left(-i\nu_n\right)+\xi_{11}}{\prod^{4}_{i=1}\left(-i\nu_{n}-\epsilon_{i{\bf{k}}_{0}{\bf{q}}}\right)},
\label{Equation_A_9}
\end{eqnarray}
\end{widetext}
where the coefficients $\xi_{i}$ ($i=1,...12$) in the expression of the polynomials ${\cal{P}}^{\left(3\right)}_{\mu}\left(i\nu_n\right)$, ${\cal{P}}^{\left(2\right)}_{\delta{\bar{n}}}\left(i\nu_n\right)$, ${\cal{P}}^{\left(2\right)}_{\Delta_{\sigma}}\left(i\nu_n\right)$ and ${\cal{P}}^{\left(2\right)}_{\Delta_{\rm s}}\left(i\nu_n\right)$ are defined as:
\begin{widetext}
\begin{eqnarray}
\xi_{1}=-6\left(\mu_{1}+\mu_{2}\right),
\nonumber\\
\xi_{2}=2\left(-2a^{2}\left(U,V\right)+8|
\Delta_{\rm s}|^{2}-2|\Delta_{\sigma}+t_{\perp}|^{2}+\mu^{2}_{1}+4\mu_{1}\mu_{2}+\mu^{2}_{2}\right),
\nonumber\\
\xi_{3}=2\left(\mu_{1}+\mu_{2}\right)\left(a^{2}\left(U,V\right)+|\Delta_{\sigma}+t_{\perp}|^{2}-4|
\Delta_{\rm s}|^{2}-\mu_{1}\mu_{2}\right),
\nonumber\\
\xi_{4}=4a\left(U,V\right)\left(\mu_{1}+\mu_{2}\right),
\nonumber\\
\xi_{5}=2a\left(U,V\right)\left(2a^{2}\left(U,V\right)+2|\Delta_{\sigma}+t_{\perp}|^{2}+8|
\Delta_{\rm s}|^{2}-\mu^{2}_{1}-\mu^{2}_{2}\right),
\nonumber\\
\xi_{6}=4\left(\Delta_{\sigma}+t_{\perp}\right),
\nonumber\\
\xi_7=-2\left(\Delta_{\sigma}+t_{\perp}\right)\left(\mu_{1}+\mu_{2}\right),
\nonumber\\
\xi_8=-2\left(\Delta_{\sigma}+t_{\perp}\right)\left(2a^{2}\left(U,V\right)+2|\Delta_{\sigma}+t_{\perp}|^{2}+8|\Delta_{\rm s}|^{2}-\mu^{2}_{1}-\mu^{2}_{2}\right),
\nonumber\\
\xi_{9}=-8\Delta_{\rm s},
\nonumber\\
\xi_{10}=8\Delta_{\rm s}\left(\mu_{1}+\mu_{2}\right),
\nonumber\\
\xi_{11}=-4\Delta_{\rm s}\left(2a^{2}\left(U,V\right)+2|\Delta_{\sigma}+t_{\perp}|^{2}+8|
\Delta_{\rm s}|^{2}+2\mu_{1}\mu_{2}\right).
\label{Equation_A_10}
\end{eqnarray}
\end{widetext}

Furthermore, we put the expressions in Eq.(\ref{Equation_A_9}) into Eq.(\ref{Equation_A_8}). 

We get:
\begin{eqnarray}
\frac{1}{\beta{N_{\rm ph}}}\sum_{{\bf{q}}\sigma}\sum^{4}_{i=1}\frac{\alpha_{i{\bf{k}}_{0}{\bf{q}}\sigma}}{-i\nu_{n}-\epsilon_{i{\bf{k}}_{0}{\bf{q}}\sigma}}=\frac{1}{\kappa},
\nonumber\\
\frac{1}{\beta{N_{\rm ph}}}\sum_{{\bf{q}}\sigma}\sum^{4}_{i=1}\frac{\beta_{i{\bf{k}}_{0}{\bf{q}}\sigma}}{-i\nu_{n}-\epsilon_{i{\bf{k}}_{0}{\bf{q}}\sigma}}=\delta{\bar{n}},
\nonumber\\
\frac{W\left(\Delta_{\sigma}+t_{\perp}\right)}{\beta{N_{\rm ph}}}\sum_{{\bf{q}}\sigma}\sum^{4}_{i=1}\frac{\gamma_{i{\bf{k}}_{0}{\bf{q}}\sigma}}{-i\nu_{n}-\epsilon_{i{\bf{k}}_{0}{\bf{q}}\sigma}}=\Delta_{\sigma},
\nonumber\\
\frac{g}{\beta{N_{\rm ph}}}\sum_{{\bf{q}}\sigma}\sum^{4}_{i=1}\frac{\delta_{i{\bf{k}}_{0}{\bf{q}}\sigma}}{-i\nu_{n}-\epsilon_{i{\bf{k}}_{0}{\bf{q}}\sigma}}=\Delta_{\rm s},
\label{Equation_A_11}
\end{eqnarray}
where the coefficients $\alpha_{i{\bf{k}}_{0}{\bf{q}}\sigma}$, $\beta_{i{\bf{k}}_{0}{\bf{q}}\sigma}$, $\gamma_{i{\bf{k}}_{0}{\bf{q}}\sigma}$ and $\delta_{i{\bf{k}}_{0}{\bf{q}}\sigma}$ have been defined as follows:
\begin{widetext}
\begin{eqnarray}
\footnotesize
\arraycolsep=0pt
\medmuskip = 0mu
\alpha_{i{\bf{k}}_{0}{\bf{q}}\sigma}
=\left\{
\begin{array}{cc}
\displaystyle  & \frac{\left(-1\right)^{i+1}}{\epsilon_{1{\bf{k}}_{0}{\bf{q}}\sigma}-\epsilon_{2{\bf{k}}_{0}\sigma}}\prod_{j=3,4}\frac{{\cal{P}}^{\left(3\right)}_{\mu}\left(\epsilon_{i{\bf{k}}_{0}{\bf{q}}\sigma}\right)}{\epsilon_{i{\bf{k}}_{0}{\bf{q}}\sigma}-\epsilon_{j{\bf{k}}_{0}\sigma}}, \ \ \ $if$ \ \ \  i=1,2.
\newline\\
\newline\\
\displaystyle  & \frac{\left(-1\right)^{i+1}}{\epsilon_{3{\bf{k}}_{0}{\bf{q}}\sigma}-\epsilon_{4{\bf{k}}_{0}\sigma}}\prod_{j=1,2}\frac{{\cal{P}}^{\left(3\right)}_{\mu}\left(\epsilon_{i{\bf{k}}_{0}{\bf{q}}\sigma}\right)}{\epsilon_{i{\bf{k}}_{0}{\bf{q}}\sigma}-\epsilon_{j{\bf{k}}_{0}{\bf{q}}\sigma}}, \ \ \ $if$ \ \ \  i=3,4,
\end{array}\right.
\label{Equation_A_12}
\end{eqnarray}
\begin{eqnarray}
\footnotesize
\arraycolsep=0pt
\medmuskip = 0mu
\beta_{i{\bf{k}}_{0}{\bf{q}}\sigma}
=\left\{
\begin{array}{cc}
\displaystyle  & \frac{\left(-1\right)^{i+1}}{\epsilon_{1{\bf{k}}_{0}{\bf{q}}\sigma}-\epsilon_{2{\bf{k}}_{0}\sigma}}\prod_{j=3,4}\frac{{\cal{P}}^{\left(3\right)}_{\delta{\bar{n}}}\left(\epsilon_{i{\bf{k}}_{0}{\bf{q}}\sigma}\right)}{\epsilon_{i{\bf{k}}_{0}{\bf{q}}\sigma}-\epsilon_{j{\bf{k}}_{0}\sigma}}, \ \ \ $if$ \ \ \  i=1,2.
\newline\\
\newline\\
\displaystyle  & \frac{\left(-1\right)^{i+1}}{\epsilon_{3{\bf{k}}_{0}{\bf{q}}\sigma}-\epsilon_{4{\bf{k}}_{0}\sigma}}\prod_{j=1,2}\frac{{\cal{P}}^{\left(3\right)}_{\delta{\bar{n}}}\left(\epsilon_{i{\bf{k}}_{0}{\bf{q}}\sigma}\right)}{\epsilon_{i{\bf{k}}_{0}{\bf{q}}\sigma}-\epsilon_{j{\bf{k}}_{0}{\bf{q}}\sigma}}, \ \ \ $if$ \ \ \  i=3,4,
\end{array}\right.
\label{Equation_A_13}
\end{eqnarray}
\begin{eqnarray}
\footnotesize
\arraycolsep=0pt
\medmuskip = 0mu
\gamma_{i{\bf{k}}_{0}{\bf{q}}\sigma}
=\left\{
\begin{array}{cc}
\displaystyle  & \frac{\left(-1\right)^{i+1}}{\epsilon_{1{\bf{k}}_{0}{\bf{q}}\sigma}-\epsilon_{2{\bf{k}}_{0}\sigma}}\prod_{j=3,4}\frac{{\cal{P}}^{\left(3\right)}_{\Delta_{\sigma}}\left(\epsilon_{i{\bf{k}}_{0}{\bf{q}}\sigma}\right)}{\epsilon_{i{\bf{k}}_{0}{\bf{q}}\sigma}-\epsilon_{j{\bf{k}}_{0}\sigma}}, \ \ \ $if$ \ \ \  i=1,2.
\newline\\
\newline\\
\displaystyle  & \frac{\left(-1\right)^{i+1}}{\epsilon_{3{\bf{k}}_{0}{\bf{q}}\sigma}-\epsilon_{4{\bf{k}}_{0}\sigma}}\prod_{j=1,2}\frac{{\cal{P}}^{\left(3\right)}_{\Delta_{\sigma}}\left(\epsilon_{i{\bf{k}}_{0}{\bf{q}}\sigma}\right)}{\epsilon_{i{\bf{k}}_{0}{\bf{q}}\sigma}-\epsilon_{j{\bf{k}}_{0}{\bf{q}}\sigma}}, \ \ \ $if$ \ \ \  i=3,4,
\end{array}\right.
\label{Equation_A_14}
\end{eqnarray}
\begin{eqnarray}
\footnotesize
\arraycolsep=0pt
\medmuskip = 0mu
\delta_{i{\bf{k}}_{0}{\bf{q}}\sigma}
=\left\{
\begin{array}{cc}
\displaystyle  & \frac{\left(-1\right)^{i+1}}{\epsilon_{1{\bf{k}}_{0}{\bf{q}}\sigma}-\epsilon_{2{\bf{k}}_{0}\sigma}}\prod_{j=3,4}\frac{{\cal{P}}^{\left(3\right)}_{\Delta_{\rm s}}\left(\epsilon_{i{\bf{k}}_{0}{\bf{q}}\sigma}\right)}{\epsilon_{i{\bf{k}}_{0}{\bf{q}}\sigma}-\epsilon_{j{\bf{k}}_{0}\sigma}}, \ \ \ $if$ \ \ \  i=1,2.
\newline\\
\newline\\
\displaystyle  & \frac{\left(-1\right)^{i+1}}{\epsilon_{3{\bf{k}}_{0}{\bf{q}}\sigma}-\epsilon_{4{\bf{k}}_{0}\sigma}}\prod_{j=1,2}\frac{{\cal{P}}^{\left(3\right)}_{\Delta_{\rm s}}\left(\epsilon_{i{\bf{k}}_{0}{\bf{q}}\sigma}\right)}{\epsilon_{i{\bf{k}}_{0}{\bf{q}}\sigma}-\epsilon_{j{\bf{k}}_{0}{\bf{q}}\sigma}}, \ \ \ $if$ \ \ \  i=3,4.
\end{array}\right.
\label{Equation_A_15}
\end{eqnarray}

\end{widetext}
%

%
%
%

\end{document}